\documentclass[twocolumn]{aastex631}

\usepackage{enumitem}
\usepackage{wrapfig}

\shorttitle{Describing VLA 1623W}
\shortauthors{Michel et al.}

\graphicspath{{./}{figures/}}

\begin{document}

\title{A millimeter-multiwavelength continuum study of VLA 1623 West}

\correspondingauthor{Arnaud Michel}
\email{arnaud.michel@queensu.ca}

\author[0000-0003-4099-9026]{Arnaud Michel}
\affiliation{Department of Physics, Engineering Physics and Astronomy, Queen’s University, Kingston, ON, K7L 3N6, Canada}

\author[0000-0001-7474-6874]{Sarah I. Sadavoy}
\affiliation{Department of Physics, Engineering Physics and Astronomy, Queen’s University, Kingston, ON, K7L 3N6, Canada}

\author[0000-0002-9209-8708]{Patrick D. Sheehan}
\affiliation{Center for Interdisciplinary Exploration and Research in Astronomy, Northwestern University, 1800 Sherman Rd., Evanston, IL 60202, USA}

\author[0000-0002-4540-6587]{Leslie W. Looney}
\affiliation{Department of Astronomy, University of Illinois, Urbana, IL 61801, USA}

\author[0000-0002-5216-8062]{Erin G. Cox}
\affiliation{Center for Interdisciplinary Exploration and Research in Astronomy, Northwestern University, 1800 Sherman Rd., Evanston, IL 60202, USA}

\begin{abstract}
VLA 1623 West is an ambiguous source that has been described as a shocked cloudlet as well as a protostellar disk. We use deep ALMA 1.3 and 0.87~millimeter observations to constrain its shape and structure to better determine its origins. We use a series of geometric models to fit the \textit{uv} visibilities at both wavelengths with \texttt{GALARIO}. Although the Real visibilities show structures similar to what has been identified as gaps and rings in protoplanetary disks, we find that a modified Flat-Topped Gaussian at high inclination provides the best fit to the observations. This fit agrees well with expectations for an optically thick highly inclined disk. Nevertheless, we find that the geometric models consistently yield positive residuals at the four corners of the disk at both wavelengths. We interpret these residuals as evidence that the disk is flared in the millimeter dust. We use a simple toy model for an edge-on flared disk and find that the residuals best match a disk with flaring that is mainly restricted to the outer disk at $R \gtrsim 30$ au. Thus, VLA 1623W may represent a young protostellar disk where the large dust grains have not yet had enough time to settle into the mid-plane. This result may have implications for how disk evolution and vertical dust settling impact the initial conditions leading to planet formation.
\end{abstract}


\section{Introduction} \label{sec:intro}

Planets form and evolve in protoplanetary disks around young stars. Recent insights from simulations and observations of young disk masses have suggested an early onset for dust growth leading to planetesimal formation in the Class 0/I protostellar disk phase \citep{Tychoniec_2020, Cridland_2022, Drazkowska_2022}. Early imprints of this process at work could explain the substructure of two possible rings and gaps at tens of au seen in the young Class I protostar, IRS 63 \citep{Segura-Cox_2020}. Investigating the morphology of young protostellar disks is therefore necessary to describe the initial conditions and subsequently the dust evolution leading to planet formation. 

The VLA 1623-2417 system is deeply embedded in a dense core within the Oph A region \citep{Pattle_2015} located at a distance of ${\sim}139$ pc \citep{Ortiz_2018, Esplin_2020}. The premier Class 0 source \citep{Andre_1990}, VLA 1623 is among the youngest protostars under the young stellar object (YSO) classification system \citep{Andre_1993,Greene_1994,Evans_2009}. It has since been identified as a hierarchical system composed of potentially four separate protostars: a very tight protobinary VLA 1623Aa and VLA 1623Ab linked to another companion VLA 1623B $1.2^{\prime\prime}$ away, as well as a VLA 1623 West located ${\sim}10.5^{\prime\prime}$ west of the triple system \citep{Looney_2000,Murillo_2013,Harris_2018,Kawabe_2018}.

The VLA 1623 Aa, Ab binary system has a large energetic outflow  \citep{Dent_1995, Yu_1997, Caratti_2006, Nakamura_2011, White_2015, Hara_2021} and is surrounded by a circumbinary disk \citep{MurilloLai_2013}. In contrast, VLA 1623B and West have been suggested to be shocked cloudlets of heated material at the edge of the outflow cavity wall \citep{Bontemps_1997, Maury_2012, Hara_2021}. However, SED analysis \citep{MurilloLai_2013}, Keplerian rotation \citep{Murillo_2013, Ohashi_2022_VLA1623}, and high-resolution dust continuum polarization observations \citep{Harris_2018, Sadavoy_2018, Sadavoy_2019} suggest these two sources are disks. This canonical protostellar system thus requires further inquiry to describe the sources in greater detail.

Here, we focus on VLA 1623 West (hereafter, VLA 1623W). Its envelope is low mass (${\sim}0.1$~M$_{\odot}$) and appears more tenuous compared to VLA 1623 Aa, Ab \citep{MurilloLai_2013, Kirk_2017}, and it does not display a clear outflow \citep{Nisini_2015,Santangelo_2015,Hara_2021}. The proper motion for VLA 1623W is consistent with it being co-moving relative to VLA 1623 B and fits an ejection scenario from VLA 1623 Aa, Ab \citep{Harris_2018} implying a common age. However, it may be a more evolved YSO (Class I) than its Class 0 companions VLA 1623 Aa, Ab, and B \citep{MurilloLai_2013}. The consistent polarization fractions and morphology from both deep ALMA 1.3~mm \citep{Sadavoy_2019} and 0.87~mm \citep{Harris_2018} observations are thought to arise from dust self-scattering in a highly inclined optically thick disk implying large ($\lambda/2\pi \sim 100-300$~$\mu$m) dust grains. As of yet, VLA 1623W has not been modeled and quantitatively described in regards to the millimeter continuum flux distribution.

To study VLA 1623W, we use geometric models to fit sensitive Stokes $I$ continuum ALMA observations at 0.87 and 1.3~mm taken from polarization data \citep{Harris_2018, Sadavoy_2019}. The models are directly fit to the \textit{uv} visibilities to constrain the source properties and test if it can be described as a typical protostellar disk. In Sect.~\ref{sec:methods}, we present the data and simple geometric models used to fit the observed ALMA \textit{uv}-visibilities. In Sect.~\ref{sec:results}, we present the results and compare how well different geometric structures model VLA 1623W. In Sect.~\ref{sec:discussion}, we propose that VLA 1623W is an optically thick flared protostellar disk and discuss the implications of a flared disk and the dangers of misinterpreting substructures in protostellar disks. Finally, we provide a summary of our findings and avenues for research in Sect.~\ref{sec:conclusion}.

\section{Methods}\label{sec:methods}

  \begin{figure*}[!ht]
    \gridline{
        \fig{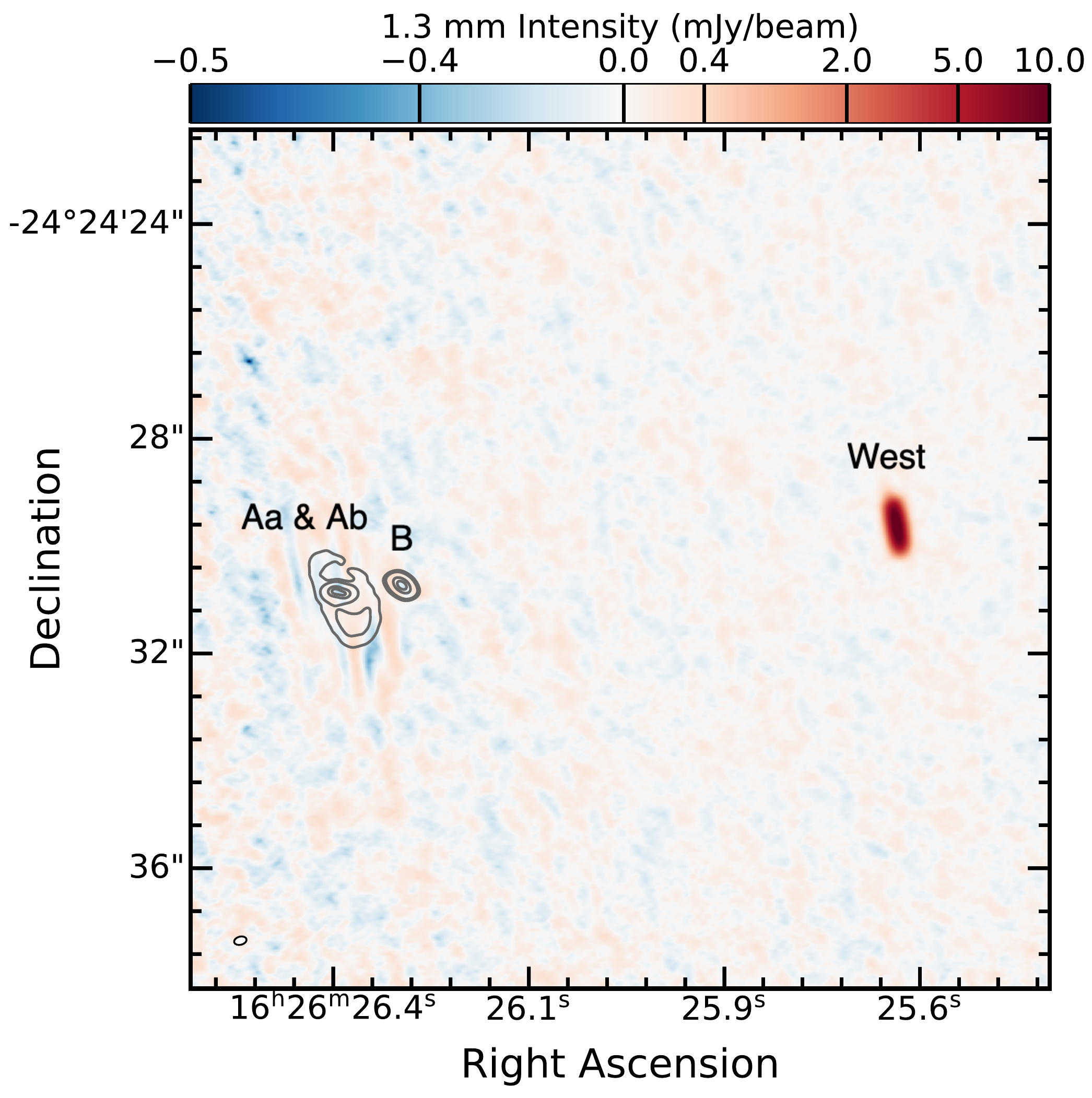}{0.365\textwidth}{(a)}
        \fig{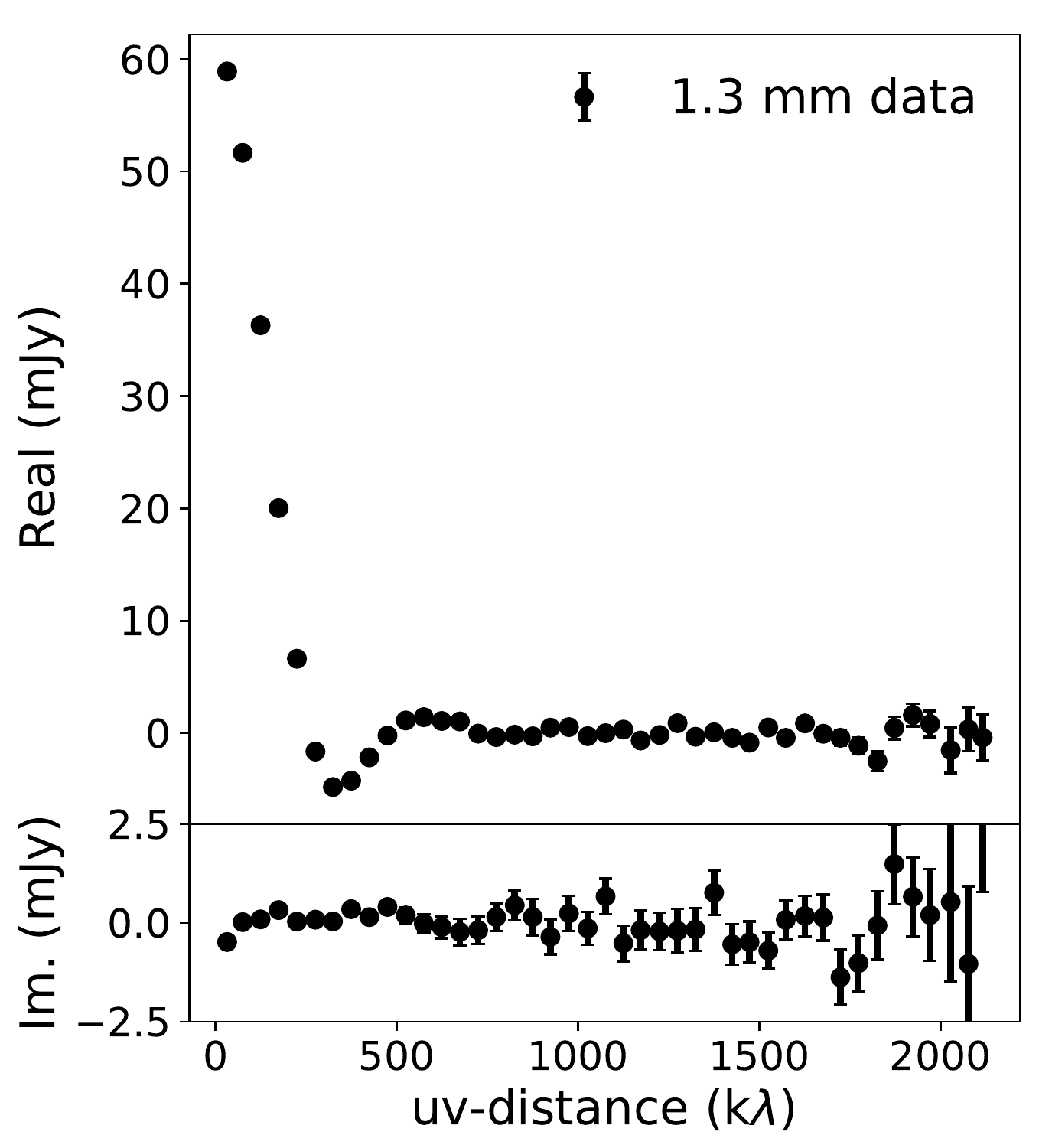}{0.317\textwidth}{(b)}
        \fig{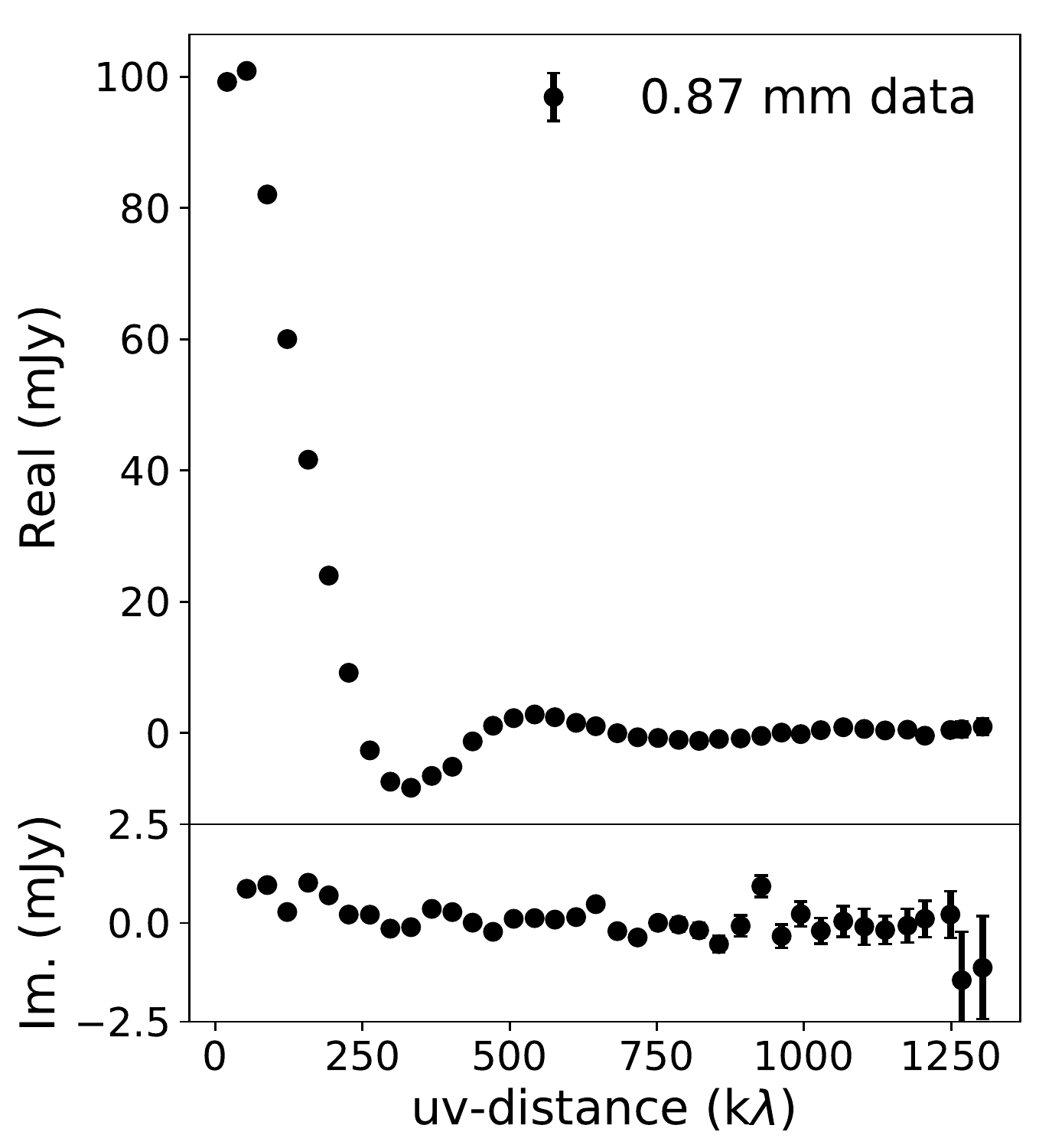}{0.317\textwidth}{(c)}
          }
    \caption{{\bf (a)} ALMA 1.3~mm field for VLA 1623W where VLA 1623 Aa, Ab \& B (left sources) have been subtracted. The residuals from Aa, Ab, \& B are on the order of $3-5\sigma_6$ while the original emission from Aa, Ab \& B, shown by the grey contours (20, 50, 100, 500$\sigma_6$), peaks at 930$\sigma_6$ (Aa \& Ab) and 980$\sigma_6$ (B). On the right, VLA 1623W, is found to peak at 220$\sigma_6$. {\bf (b)} ALMA 1.3~mm deprojected and Aa, Ab, B-subtracted \textit{uv} visibilities centered on VLA 1623W using an inclination of $80.3^{\circ}$ and a Position Angle of $10.3^{\circ}$. The \textit{uv} visibility profiles always display the full data for completeness. At 1.3~mm only the first binned data point is below 50~k$\lambda$, located at 40~k$\lambda$. {\bf (c)} Same as panel (b) but for 0.87~mm data. At 0.87~mm only the first binned data point is below 50~k$\lambda$, it is located at 20~k$\lambda$. \label{fig:AB_removal_vis}}
 \end{figure*}
 
\subsection{ALMA observations}\label{sec:almaobs}
 
VLA 1623W was observed by ALMA polarization projects at 1.3 mm (Band 6), program 2015.1.01112.S (PI: S. Sadavoy), and at 0.87 mm (Band 7), program 2015.1.00084.S (PI: L. Looney). Time on source totaled 7.5~mins at Band 6 with a $0.27 \times 0.21^{\prime\prime}$ beam \citep{Sadavoy_2019} and 80~mins at Band 7 with a $0.17 \times 0.15^{\prime\prime}$ beam \citep{Harris_2018}. Both data-sets report polarization fractions ${\sim}1-1.5\%$ for VLA 1623W and a uniform polarization aligned with the minor axis. The deep observations required for polarization detections yield excellent Stokes I continuum sensitivities with which we can detect faint structures \citep{Gulick_2021}.

Both sets of ALMA observations include VLA 1623 Aa, Ab, B, and West within the primary beam. While the 1.3~mm observations are centered on West, the 0.87~mm data are centered at an equidistant point between the triple system (Aa, Ab, B) and West. We remove VLA 1623 Aa, Ab, and B by modeling these with \texttt{TCLEAN} using Briggs weighting and a robust parameter of 0.5 in \texttt{CASA 5.2.6.1}. The Fourier transform of the \texttt{TCLEAN} modeled emission from VLA 1623 Aa, Ab, and B is subtracted from the field using \texttt{uvsub} such that VLA 1623W is the only remaining source of dominant emission, see Figure~\ref{fig:AB_removal_vis}.
 
To further limit confusion in the visibilities from unrelated extended background emission in the residuals, we apply a 50~k$\lambda$ cut to both data-sets. This has been shown to remove some extended emission and improve the disk's rms sensitivity \citep{Eisner_2018,Boyden_2020}. This \textit{uv} cut corresponds to removing scales ${\gtrsim}550$ au and will not significantly impact VLA 1623W itself which has a compact size of ${\sim}100$ au \citep{Sadavoy_2019}.

The data above 50~k$\lambda$ are imaged using robust=0.5 in \texttt{CASA}. The primary beam corrected sensitivities and the beam sizes for the two wavelengths are $\sigma_6=54\mu$Jy/beam and ($0.24 \times 0.15^{\prime\prime}$) at 1.3~mm (Band 6) and $\sigma_7 = 110\mu$Jy/beam and ($0.16 \times 0.15^{\prime\prime}$) at 0.87~mm (Band 7). Figure~\ref{fig:AB_removal_vis} (a) shows VLA 1623W and the successful removal of VLA 1623 Aa, Ab, and B at 1.3~mm. The residuals from VLA 1623 Aa, Ab, and B are on the order of 3-5$\sigma_6$ at 1.3~mm and 3-5$\sigma_7$ at 0.87~mm.

The visibilities weights provided by the ALMA pipeline during the calibration process have been suggested to be too large and overconfident thus requiring re-scaling \citep[e.g.,][]{Sheehan_2020}. Following \citet{Sheehan_2020} we compare the root mean square of the naturally weighted \texttt{TCLEAN} image ($\sigma_{vis}$) to the uncertainties from the visibility weights ($w_i$, weight of each $i^{\text{th}}$ visibility), which should follow $\sigma_{vis} = \sqrt{1 / \Sigma w_i}$. We find a necessary scaling factor of 0.25 for the visibility weights for both the 1.3 and 0.87~mm data, consistent with the scale factor found by \citet{Sheehan_2020}, i.e. increasing the theoretical noise factor by a factor of 2.

\subsection{Geometric models}\label{sec:models}

We aim to characterize VLA 1623W's disk morphology with the available ALMA observations. Figure~\ref{fig:AB_removal_vis} presents the azimuthally averaged \textit{uv} visibilities centered on West. The Real \textit{uv} profiles show intensity dips below zero at ${\sim330}k\lambda$ and a slight positive enhancement at ${\sim}550k\lambda$ at both wavelengths. The Imaginary (Im) profiles, however, show scatter but little obvious structure. As such, we adopt axis-symmetric geometric models to describe the intensity profiles of this source. 

Assuming VLA 1623W is a disk, we model it in the visibility plane using a variety of analytic profiles that have been applied to protostellar and protoplanetary disks \citep[e.g., see][for a brief list]{Tazzari_multi_2021}. For simplicity, we focus on models using standard Gaussians and modified Gaussians with flat tops. For the standard Gaussian model, we included additional Gaussian features in the fitting as dips and peaks in Real \textit{uv} visibilities are often indicative of gaps and rings \citep[e.g.,][]{Andrews_2021}. We use negative Gaussians and positive Gaussians to represent these features respectively.

We also test a modified \textit{Flat-Topped Gaussian} (\textit{FTG}) model. This model is motivated by the possibility that VLA 1623W as a highly inclined and optically thick protostellar disk, as suggested from polarization observations \citep{Harris_2018,Sadavoy_2019}. ALMA millimeter observations of edge-on optically thick disks have found flat brightness profiles along the major axis \citep{Villenave_2020}. Furthermore, for such sources, the disk edge can result in a sharp drop in the millimeter emission \citep{Villenave_2020,Miotello_2022} that cannot be well represented by a regular Gaussian taper. Therefore in the \textit{FTG}, we use a Gaussian function whose exponent, $\phi$ is left as a free parameter $\geq2$ as,
\begin{equation}\label{eq:modgauss}
    I(R) = I_0 \text{exp}\left(-0.5\left(\frac{R}{\sigma}\right)^{\phi}\right),
\end{equation}
where $I(R)$ is the intensity as a function of radius $R$, $I_0$ is the peak intensity at the center, and $\sigma$ is the standard deviation width. In the case of the standard Gaussian disk and for the gap and ring features, we fix $\phi=2$.

Lastly, we test a power-law core with an exponential tail (\textit{PLCT}) \citep[e.g.,][]{Lynden-Bell_1974, Andrews_2009, Segura-Cox_2020}, which has been used to describe protostellar and protoplanetary disks. The \textit{PLCT} model can characterize both the disk and the inner envelope components thanks to its two-part surface density structure \citep{Andrews_2009}, making it a valuable tool for protostellar sources with envelope emission traced by millimeter observations. However, at both wavelengths, the parameter regulating the envelope contribution goes to 0 effectively making a standard Gaussian model for VLA 1623W. Therefore, we will not discuss the results of the \textit{PLCT} model further.

\begin{table*}[!t]
\begin{center}
\caption{Disk profiles}
\label{tbl:bestfits}
\begin{tabular}{l|lll|c}
\hline
\hline
& \multicolumn{3}{c|}{Best-fit Parameters} & Prior \\
Profile & $\lambda$ & 1.3~mm & 0.87~mm & \\
\hline

Gaussian Disk & $F_D$ (mJy) & $66^{+6}_{-6}$ & $117^{+1}_{-1}$  & $-4<\log_{10}F_D <0$\\
& $\sigma_D$ (mas) & $300^{+1}_{-1}$ & $317^{+1}_{-1}$ & $10 < \sigma_D < 2000$\\
& Maximum residuals & 23.4$\sigma_6$ & 23.4$\sigma_7$ & \\
\hline

Gaussian Disk, Gap & $F_D$ (mJy) & $161^{+16}_{-18}$ & $325^{+18}_{-11}$  & $-4<\log_{10}F_D <0$\\
& $\sigma_D$ (mas) & $215^{+4}_{-3}$ & $222^{+7}_{-9}$ & $10 < \sigma_D < 2000$\\
& $F_G$ (mJy) & $-97^{+17}_{-18}$ & $-214^{+27}_{-17}$ & $ -4<\log_{10}F_G<-1$ \\
& $\sigma_G$ (mas) & $150^{+5}_{-7}$ & $169^{+2}_{-1}$ & $1<\sigma_G<500$\\
& loc$_G$ (mas) & $61^{+9}_{-6}$ & $46^{+1}_{-2}$ & $10<\text{loc}_G<200$ \\
& Maximum residuals & 7.2$\sigma_6$ & 13.4$\sigma_7$ & \\
\hline

Gaussian Disk, Gap, Ring & $F_D$ (mJy) & $60^{+6}_{-7}$ & $106^{+1}_{-1}$  & $-4<\log_{10}F_D <0$\\
 & $\sigma_D$ (mas) & $274^{+1}_{-1}$ &  $274^{+1}_{-1}$ & $10 < \sigma_D < 2000$\\
 & $F_G$ (mJy) & $-22^{+6}_{-9}$ & $-4^{+7}_{-8}$& $ -4<\log_{10}F_G<-1$\\
& $\sigma_G$ (mas) & $238^{+19}_{-52}$ & $2^{+1}_{-1}$ & $ 1<\sigma_G<500$\\
& loc$_G$ (mas) & $139^{+50}_{-98}$ & $158^{+1}_{-1}$ & $10<\text{loc}_G<200$  \\
& $F_R$ (mJy) & $25^{+2}_{-4}$ & $10^{+3}_{-3}$ & $-4<\log_{10}F_R<-1$ \\
& $\sigma_R$ (mas) & $97^{+6}_{-8}$ & $2^{+1}_{-1}$ & $1<\sigma_R<500$ \\
& loc$_R$ (mas) & $359^{+4}_{-6}$ & $432^{+1}_{-1}$ & $200<\text{loc}_R<500$ \\
& Maximum residuals & 6.5$\sigma_6$ & 11.6$\sigma_7$ & \\
\hline

Flat-Topped Gaussian & $F_D$ (mJy) & $63^{+1}_{-1}$ & $110^{+1}_{-1}$ & $-4<\log_{10}F_D <0$\\
& $\sigma_D$ (mas) & $449^{+1}_{-1}$ & $459^{+1}_{-1}$ & $10 < \sigma_D < 2000$\\
& $\phi$ & $5.01^{+0.07}_{-0.07}$ & $4.95^{+0.03}_{-0.03}$ & $ 2<\phi<10$ \\
& Maximum residuals & 5.8$\sigma_6$ & 13.3$\sigma_7$ & \\
\hline

\end{tabular}
\end{center}
\vspace{-1mm}
\textit{Notes:}
\begin{enumerate}[noitemsep]
    \item $F_X$ is the total integrated flux of each function, Gaussian disk ($F_D$), Gaussian gap ($F_G$), and Gaussian ring ($F_R$).
    \item The Gaussian substructures, gaps and rings are offset from the center by a distance loc$_X$.
    \item Maximum residuals are positive excess emission. In all cases, the positive residuals are greater than the negative ones. The residuals are obtained from the cleaned image of the observed visibilities minus the best-fit model visibilities.
    \item At 0.87~mm for the \textit{Gaussian Disk, Gap, Ring} model, the value of $\sigma_G$ over multiple runs appears bimodal, either converging to $\sigma_G=2$~mas (with loc$_G\approx150$~mas) or $\sigma_G=1$0~mas (loc$_G\approx53$~mas). The BIC condition (see Section~\ref{sec:stats}) strongly favours the first solution, so we only consider those parameters further.  The remaining parameters are consistent between both cases.
\end{enumerate}
\end{table*}

For each disk model, we fit the model free parameters as well as the inclination $i$ of the disk along the line of sight, the position angle $PA$, and the source offsets from the field center $\Delta$R.A. and $\Delta$Dec (see Table~\ref{tbl:bestfits}). Note that we fit the total integrated flux ($F$) as a free parameter rather than the peak surface brightness ($I_0$) similar to \citet{Sheehan_2020}. The parameter space thus ranges from being 6-dimensional (Gaussian disk) to 12-dimensional (Gaussian disk, gap, ring). We use uniform priors for all parameters, see Table~\ref{tbl:bestfits} for the ranges. We explore the parameter space for the models with a Bayesian approach using an affine-invariant Markov chain Monte Carlo (MCMC) ensemble sampler, \texttt{emcee} v 2.2.1 \citep{Foreman-MacKey_2013}, within the \texttt{GALARIO} python package \citep{Tazzari_2018}. 

We generate a $5^{\prime\prime}$ radial grid with a $10^{-5}$ arcsecond cell size for which the geometric model is evaluated. \texttt{GALARIO} then allows us to Fourier transform the 2D geometric models to synthetic visibilities using the baseline pairs from the observations, thereby generating synthetic observations of the model on the same scale as the true observations. We minimize the $\chi^2$ value between the sum of the observed Real and Imaginary visibilities compared to the synthetic equivalent to find the optimal fit parameters using 60 walkers over 5000 steps distributed using \texttt{MPIPool}. We image the residuals in \texttt{CASA}, which we obtain by subtracting the Fourier transform of the disk models from the observations in \texttt{GALARIO}, to examine the quality of the fit in the image plane. 

\section{Results and Analysis}\label{sec:results}

\subsection{Model Fitting Results}\label{sec:model_fit_res}

 \begin{figure*}[!ht]
 \centering
 \textbf{\underline{Gaussian disk model}}
    \gridline{\fig{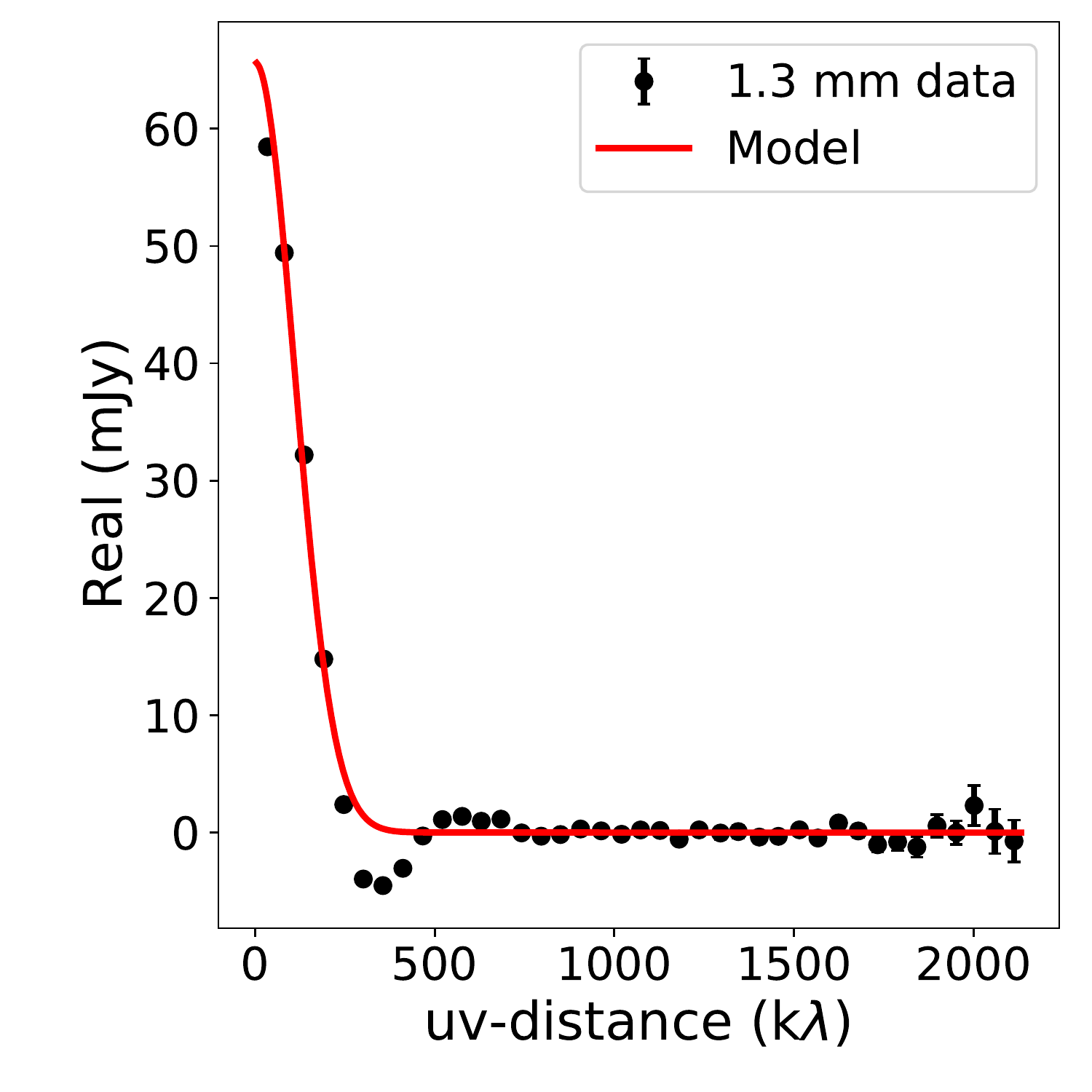}{0.245\textwidth}{}
          \fig{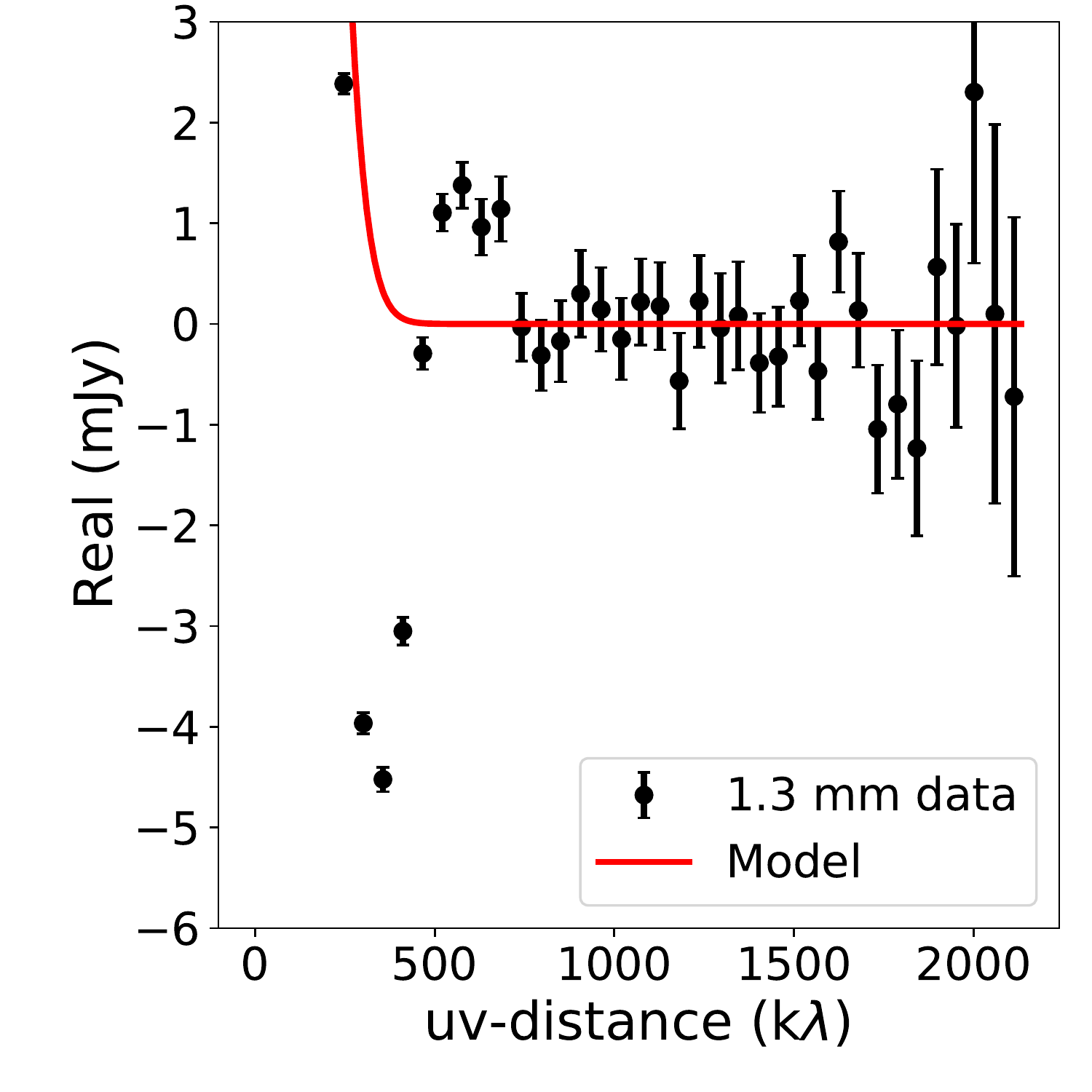}{0.245\textwidth}{}
          \fig{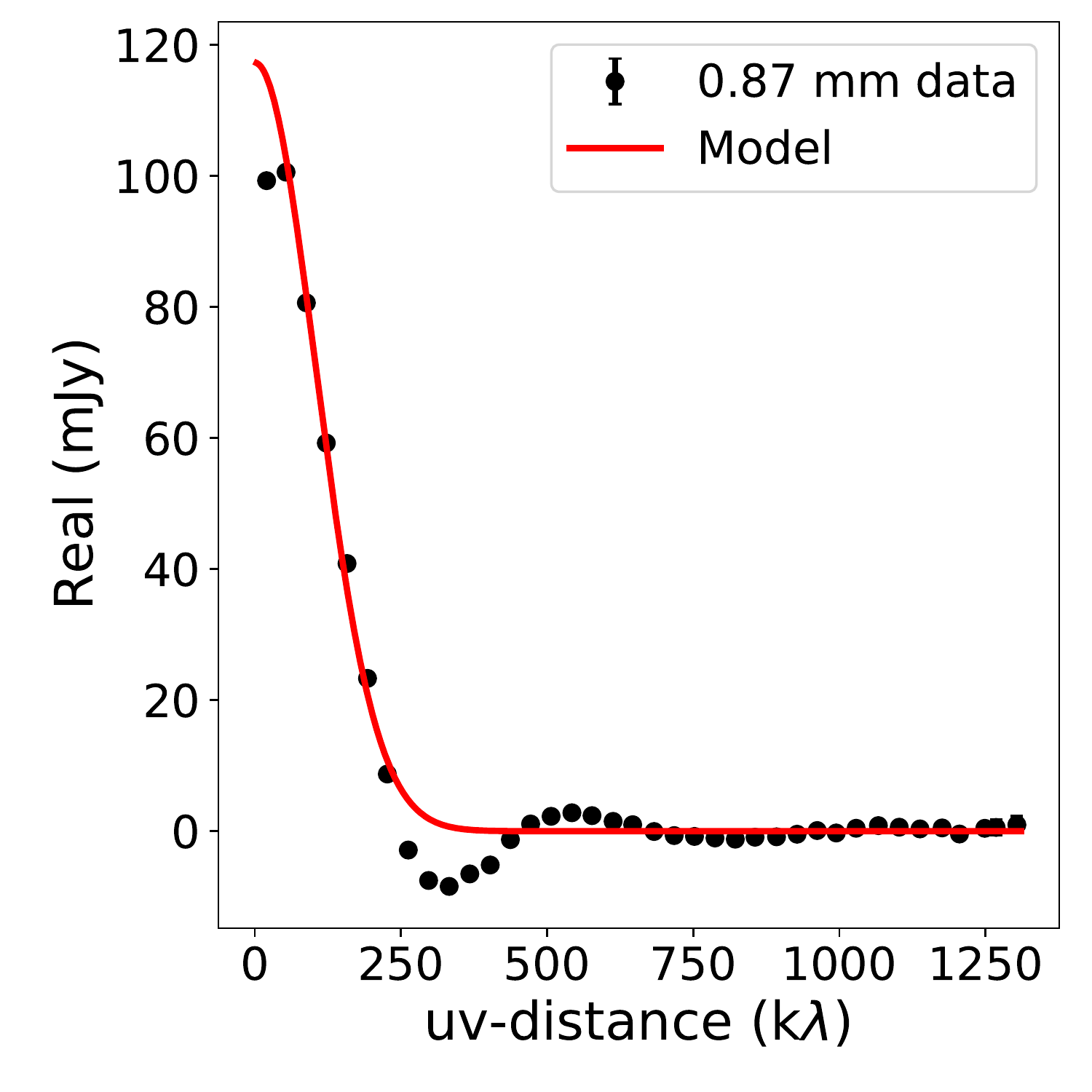}{0.245\textwidth}{}
          \fig{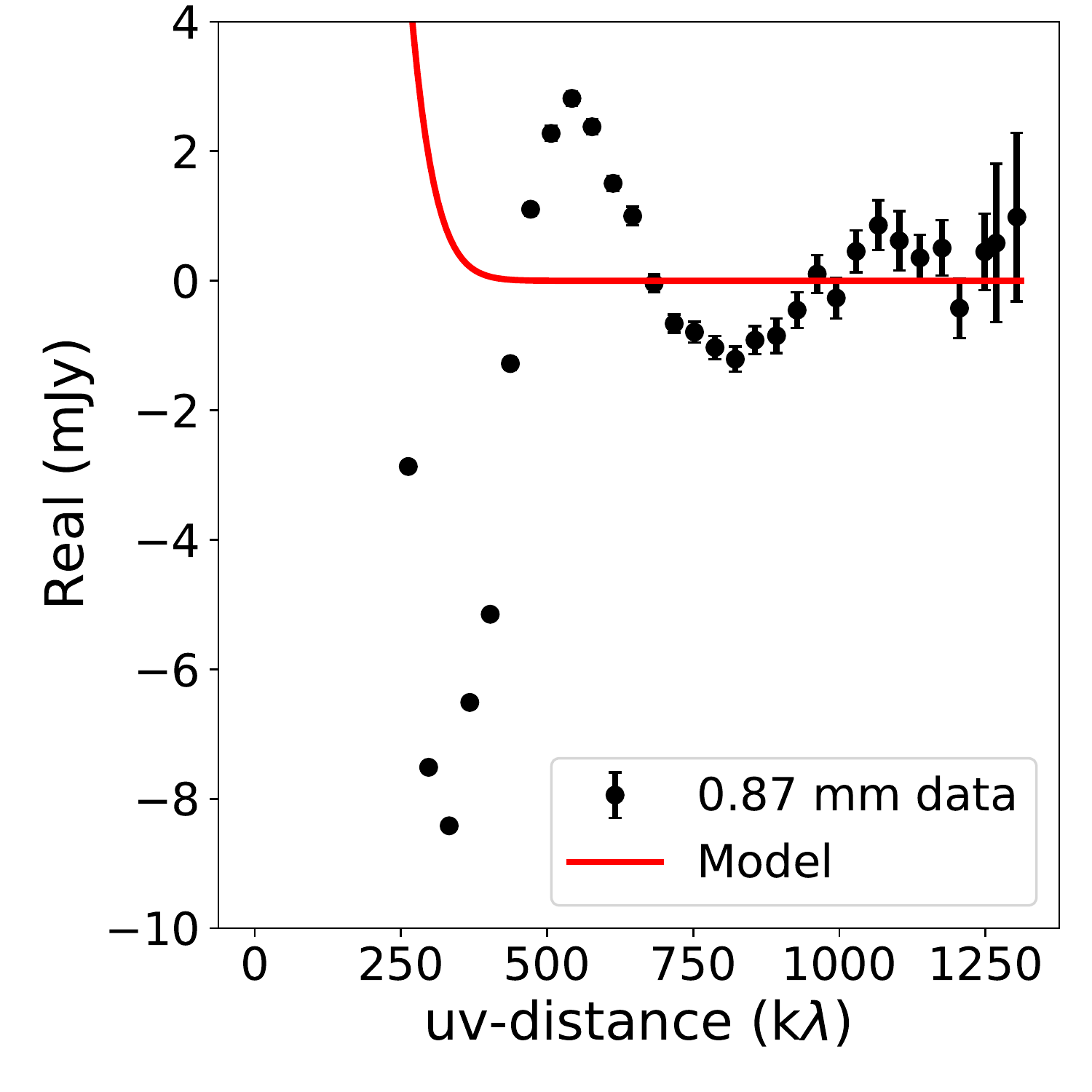}{0.245\textwidth}{}}
    \vspace{-4mm}      
    \textbf{\underline{Gaussian disk, gap model}}
    \gridline{\fig{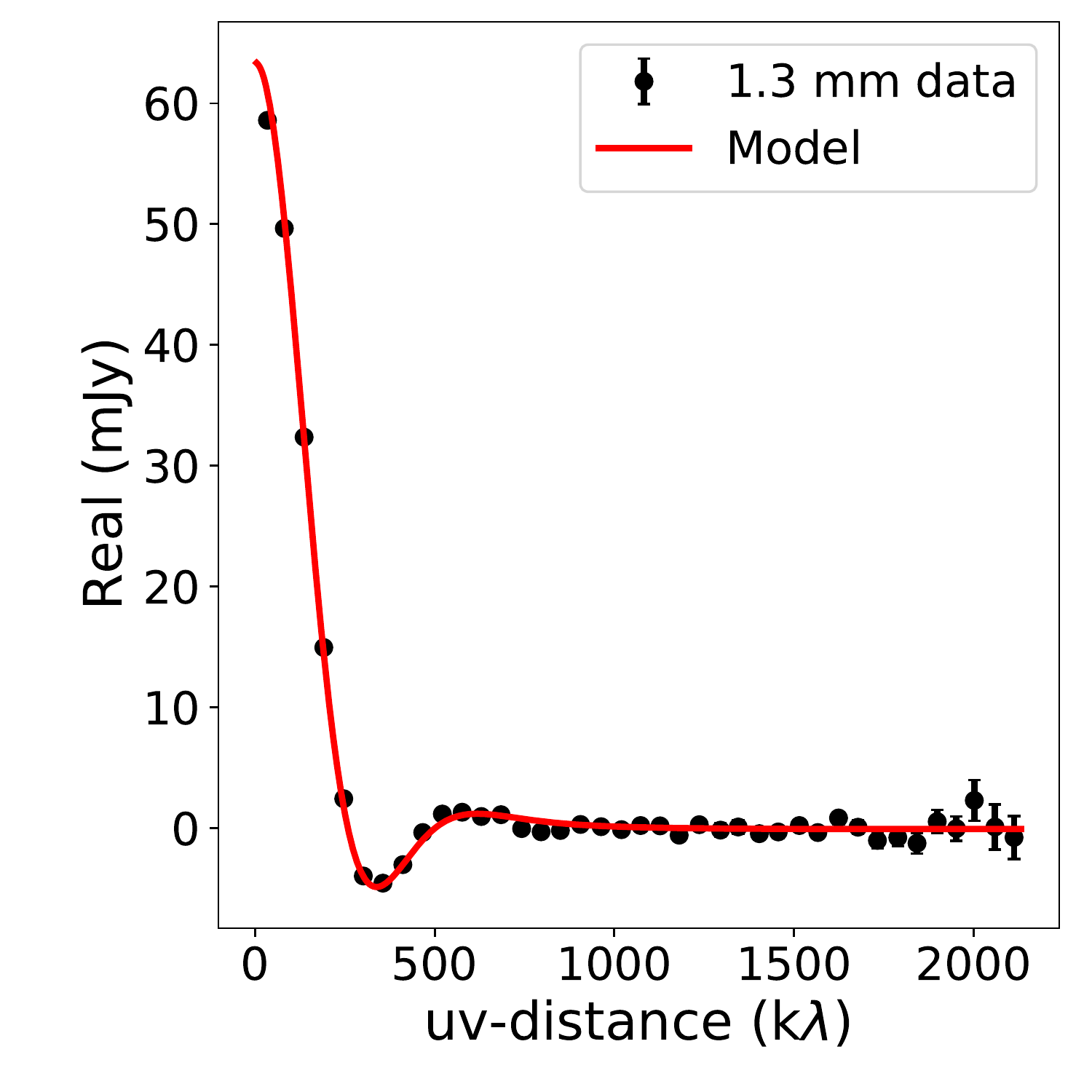}{0.245\textwidth}{}
          \fig{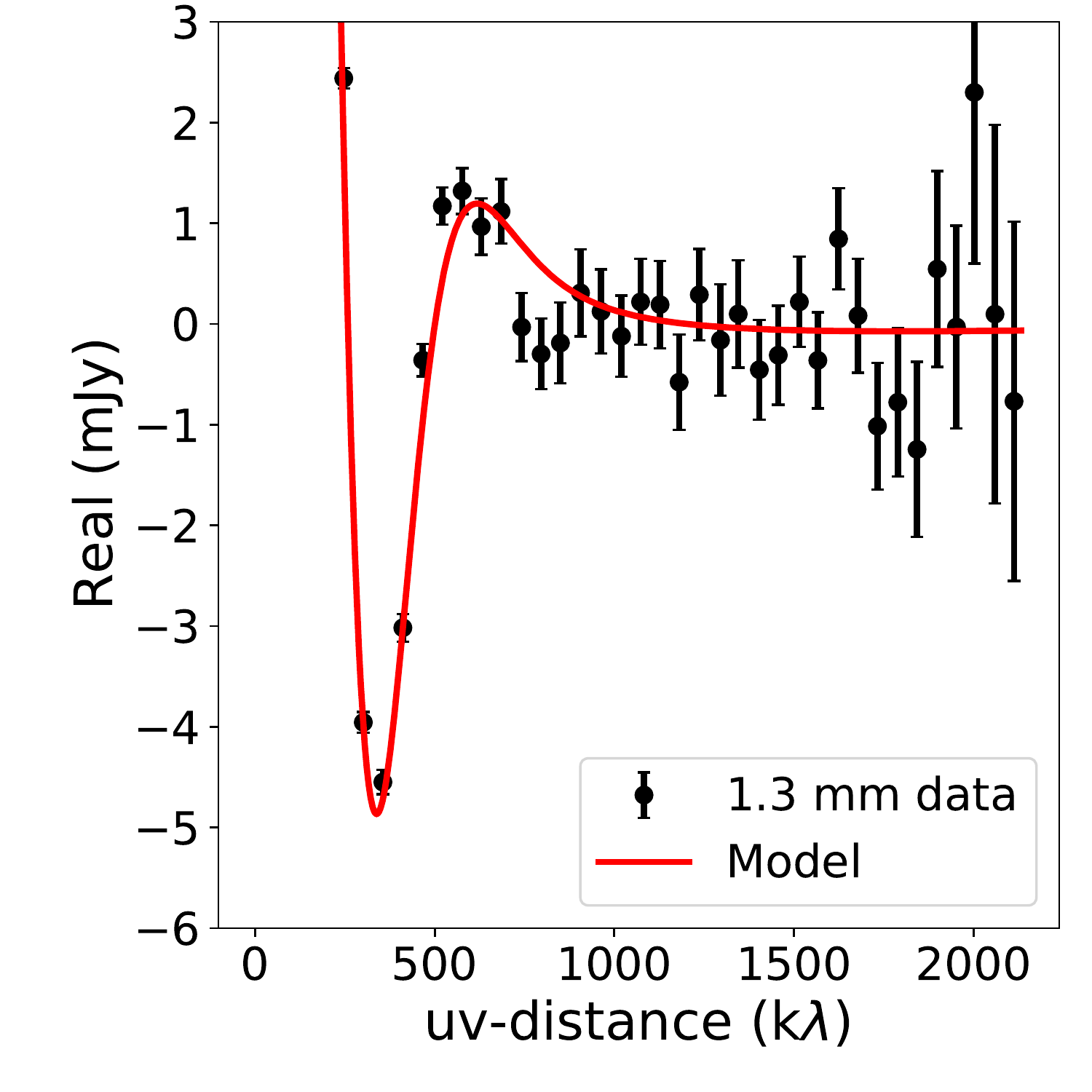}{0.245\textwidth}{}
          \fig{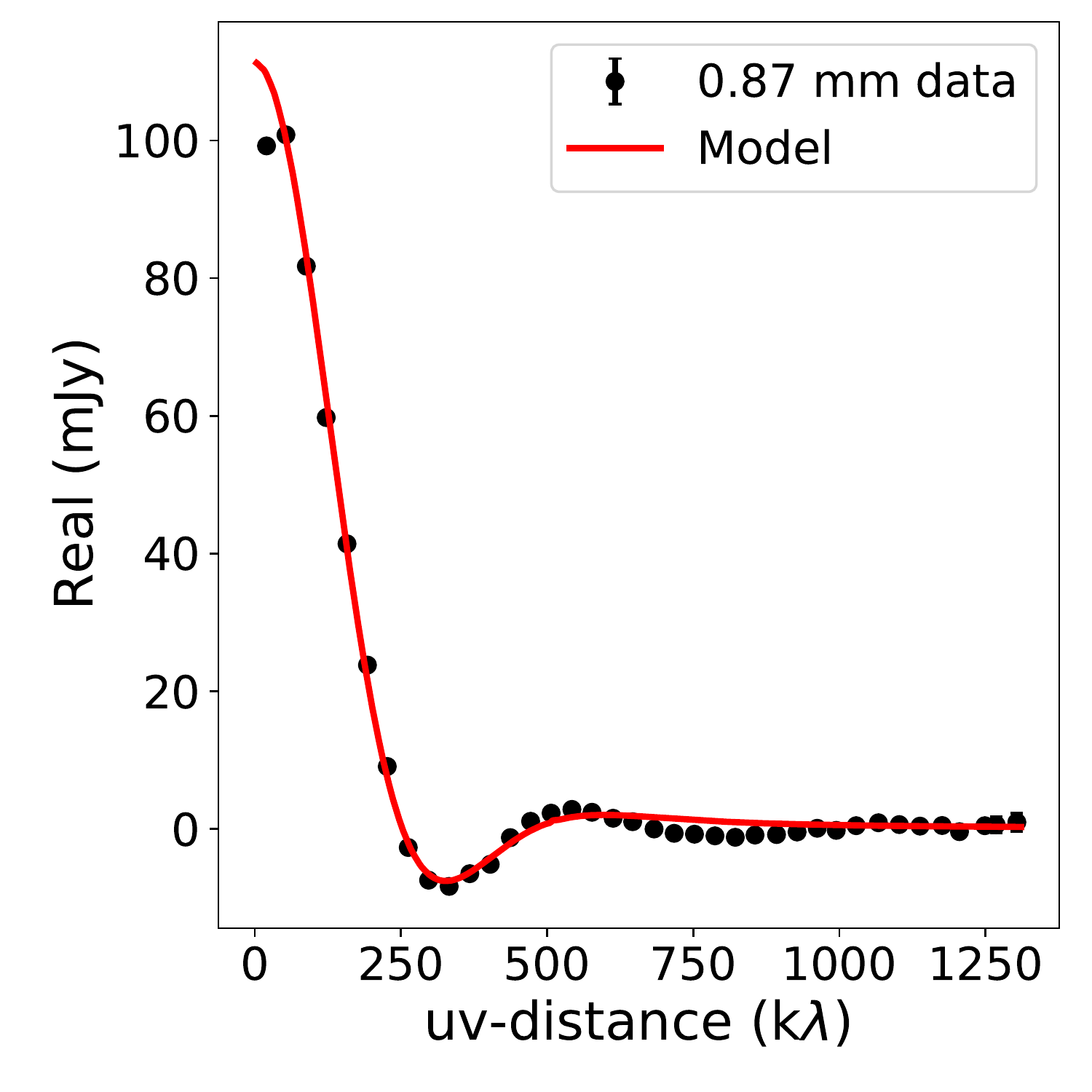}{0.245\textwidth}{}
          \fig{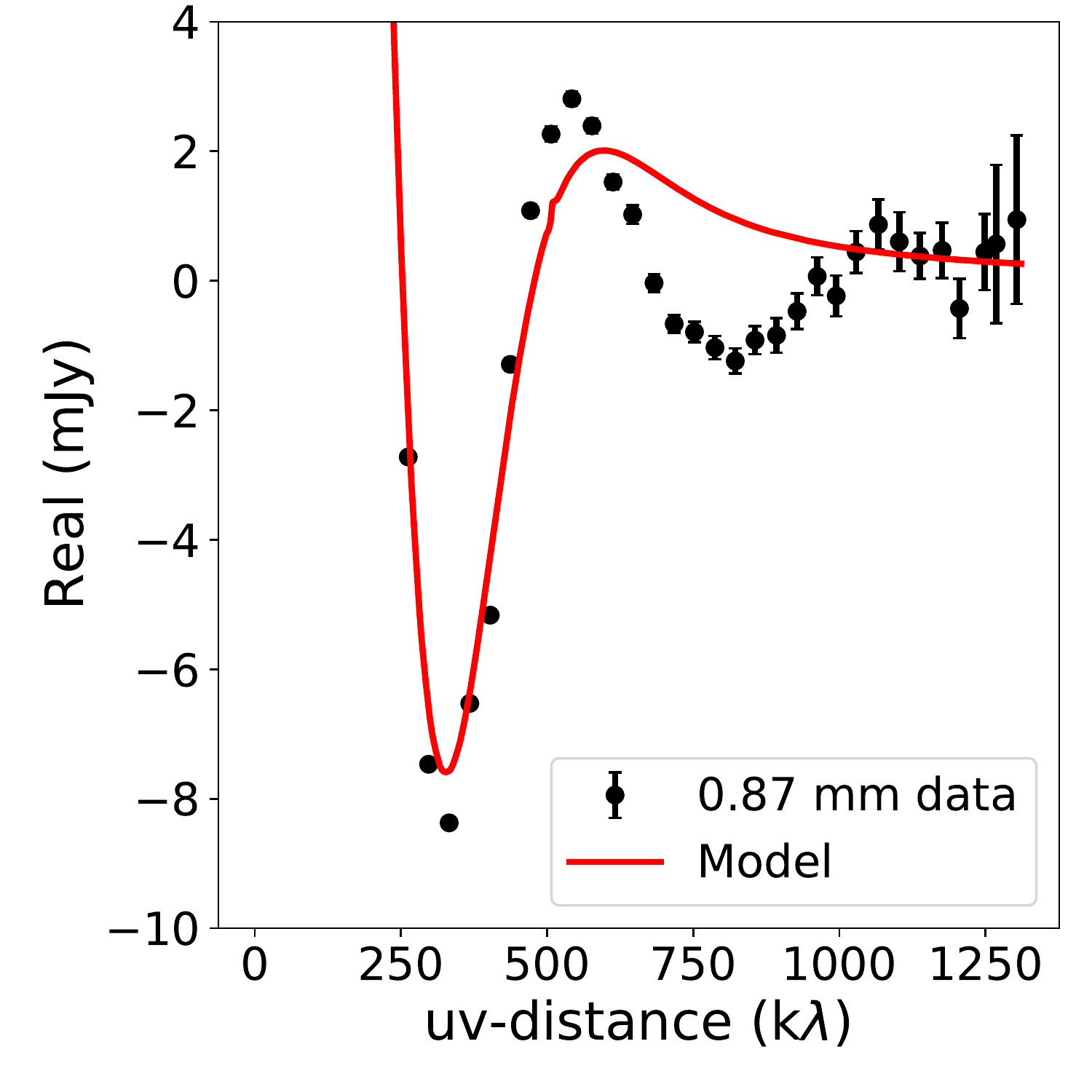}{0.245\textwidth}{}}
    \vspace{-4mm}       
    \textbf{\underline{Gaussian disk, gap, ring model}}
    \gridline{\fig{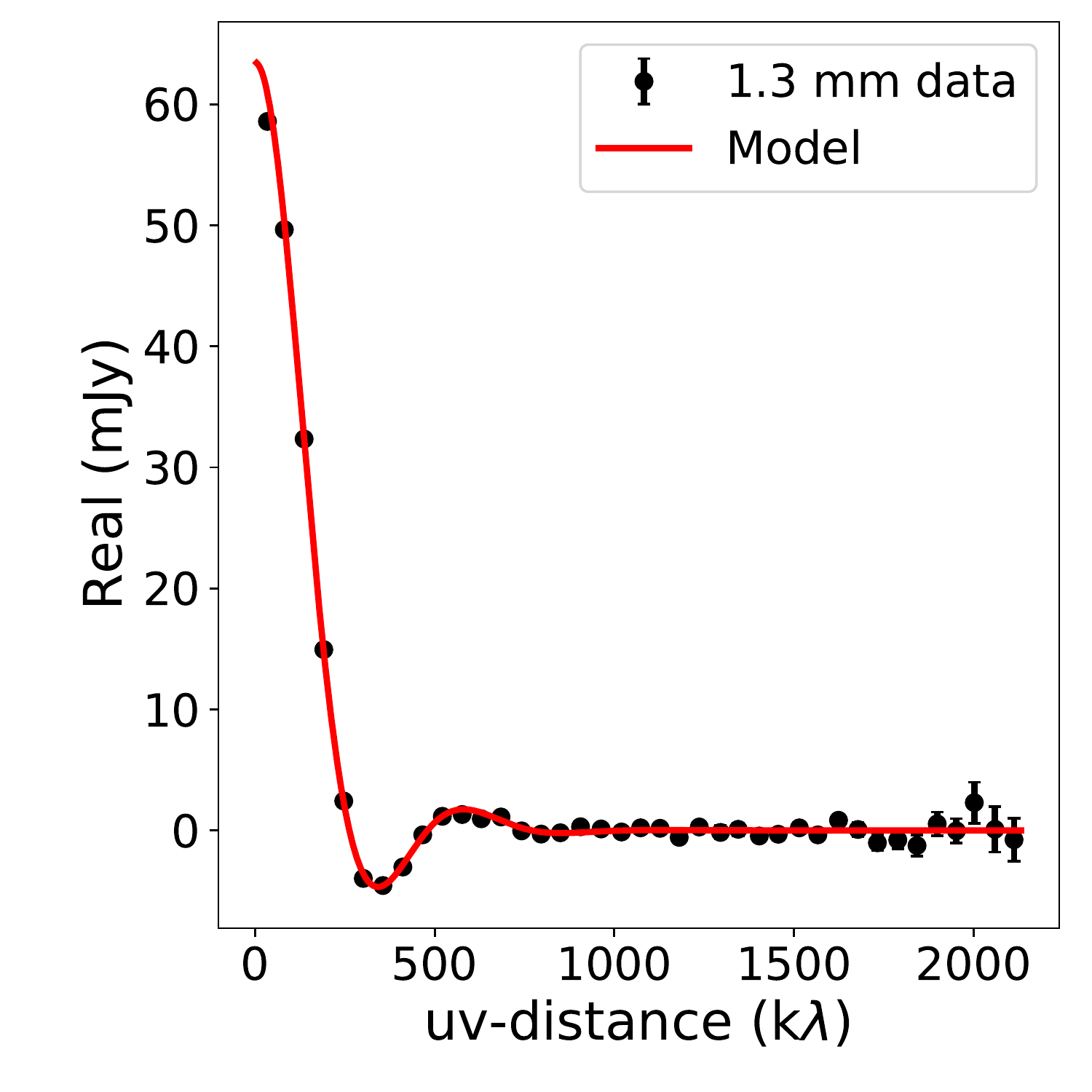}{0.245\textwidth}{}
          \fig{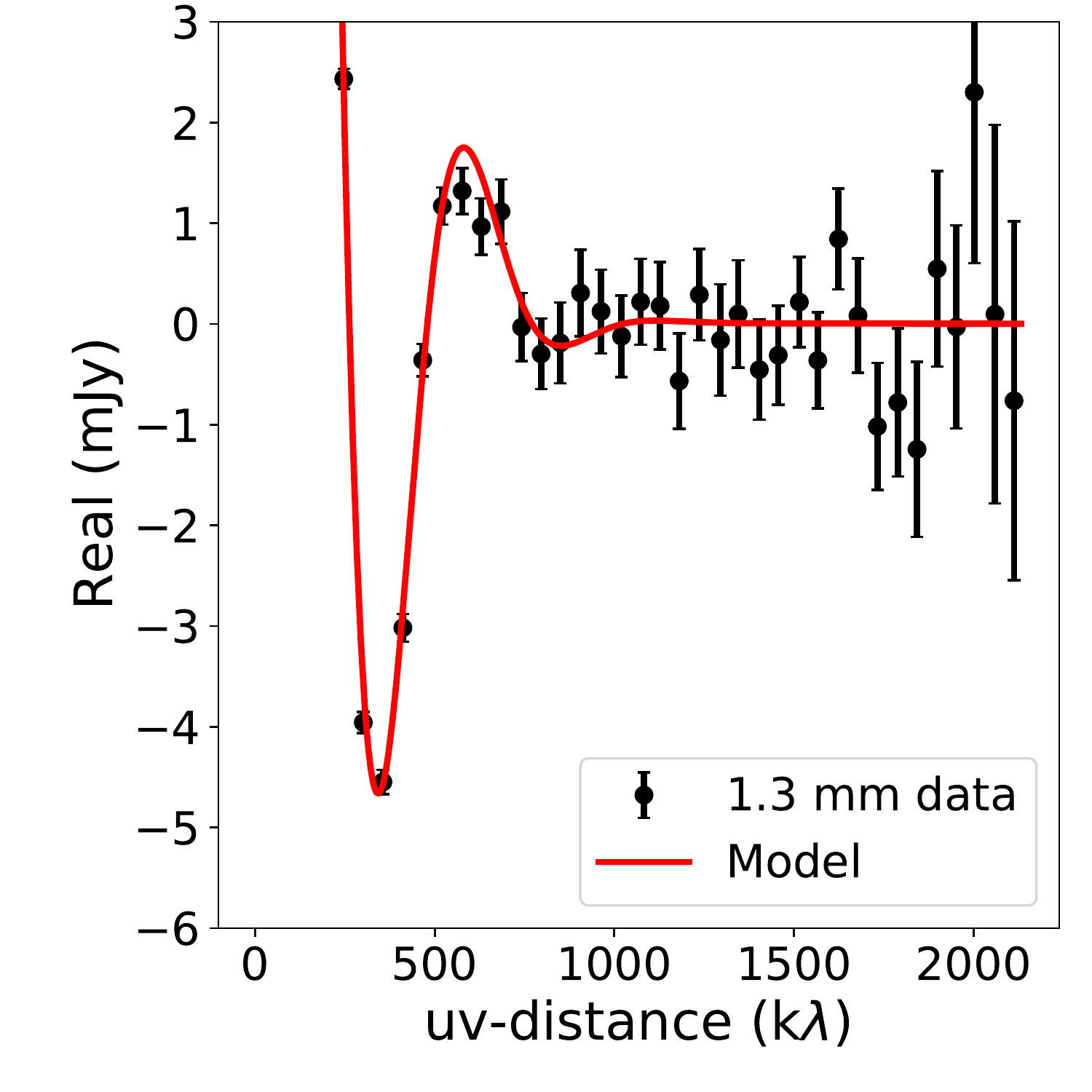}{0.245\textwidth}{}
          \fig{VLA1623W_B7_gauss_gap_ring_iPAoffsets_min50k_galario_noIM.pdf}{0.245\textwidth}{}
          \fig{VLA1623W_B7_gauss_gap_ring_iPAoffsets_min50k_galario_noIM_cut.pdf}{0.245\textwidth}{}}
    \vspace{-4mm}       
    \textbf{\underline{Flat-Topped Gaussian model}}
    \gridline{\fig{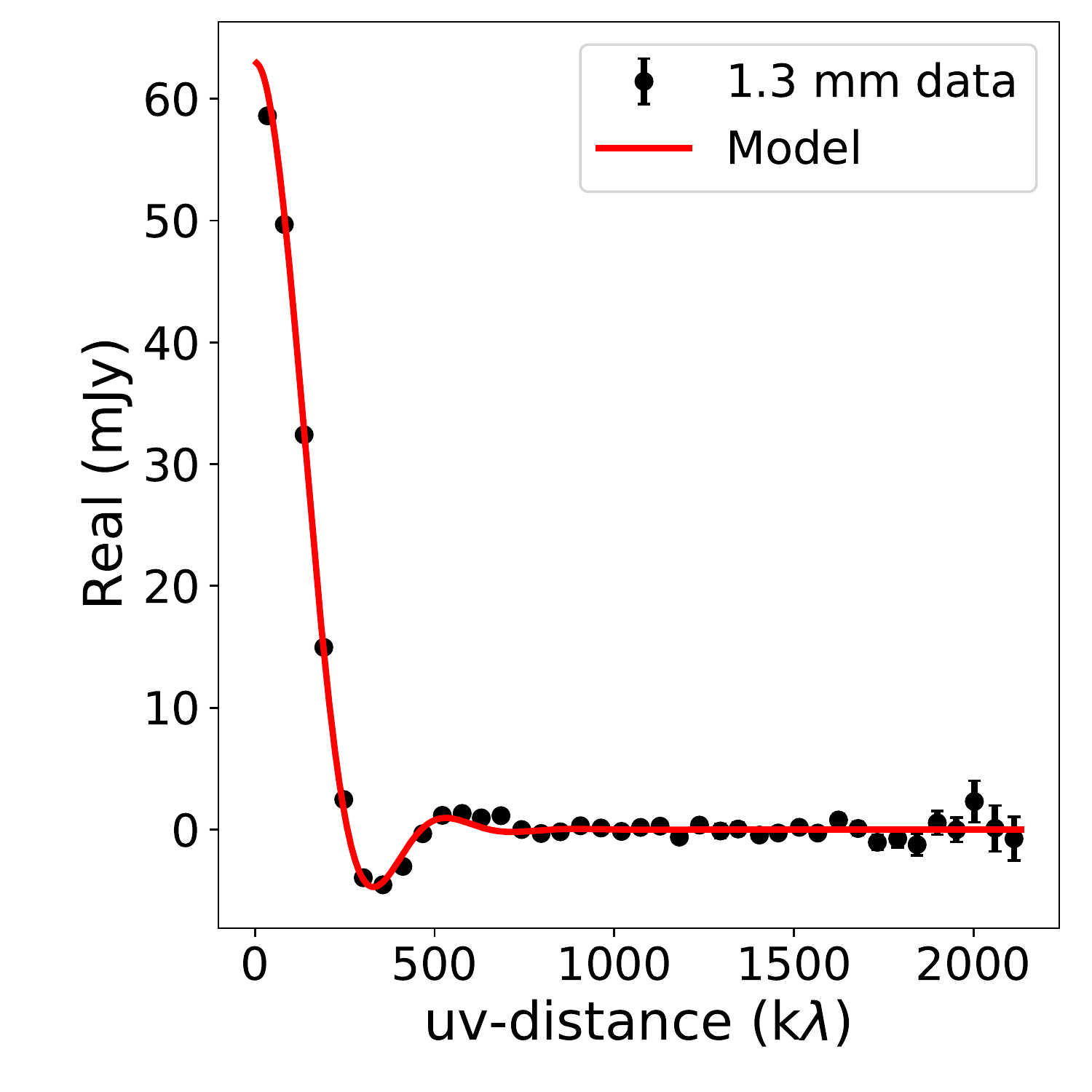}{0.245\textwidth}{}
          \fig{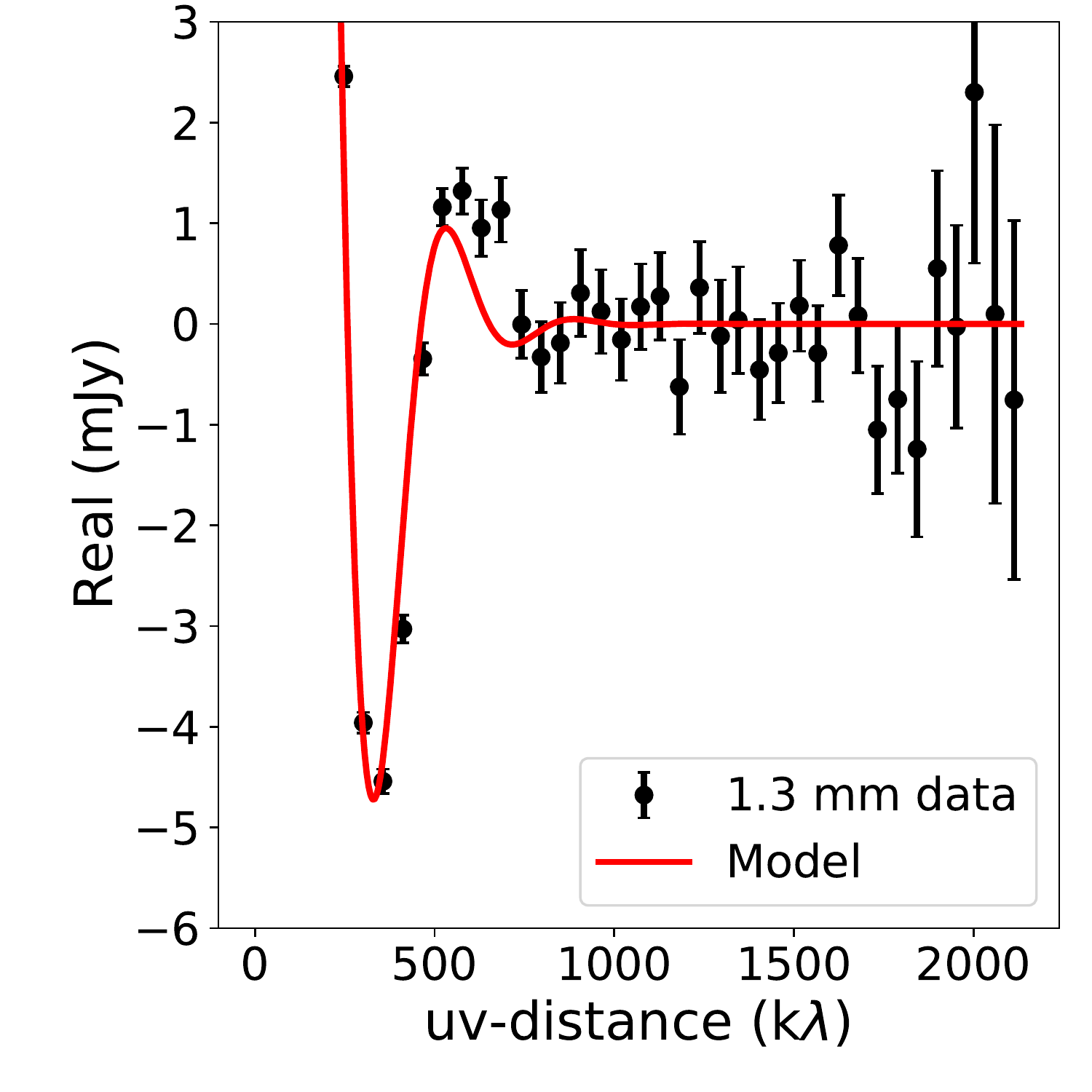}{0.245\textwidth}{}
          \fig{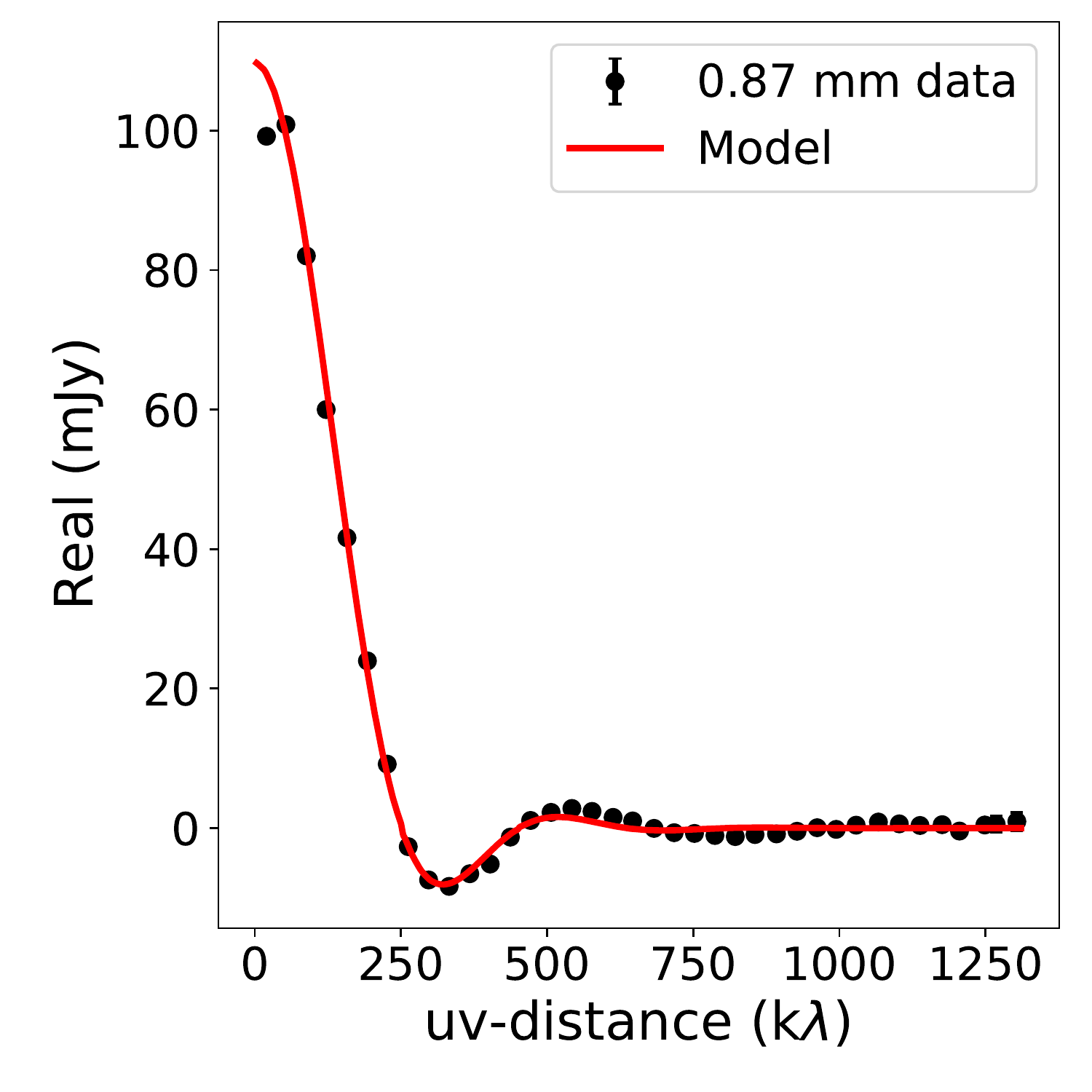}{0.245\textwidth}{}
          \fig{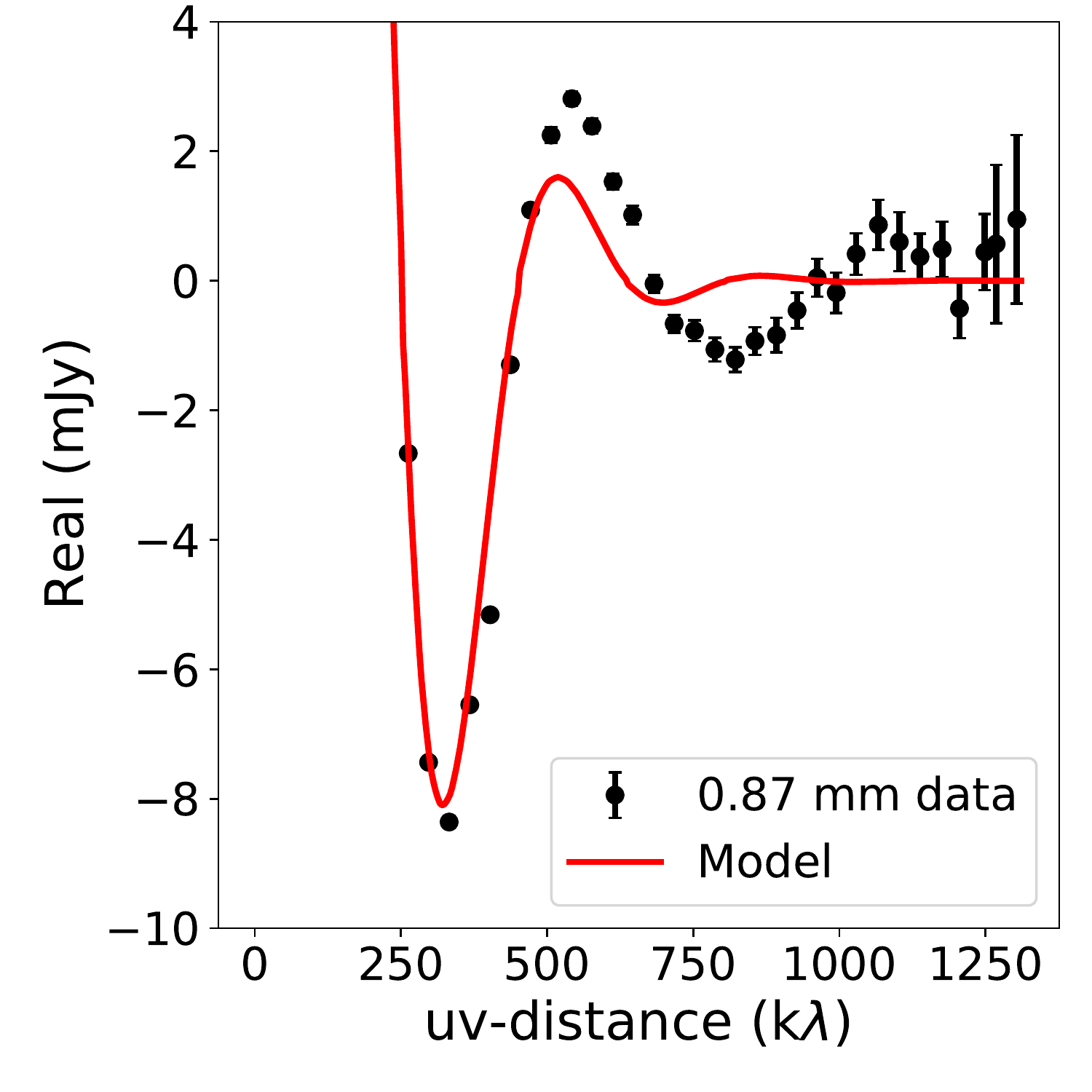}{0.245\textwidth}{}}
    \vspace{-5mm} 
    \caption{Real \textit{uv} visibility profiles in black data points fit by a series of different disk models described in Section~\ref{sec:models} shown by the red lines, see Table~\ref{tbl:bestfits} for the best-fit parameters shown here. The first left column is the full 1.3~mm data while the second column is a zoom in to highlight the region of structured visibilities. The third and fourth columns display the same information for the 0.87~mm observations. The topmost row shows the standard \textit{Gaussian disk} model, the second row shows the \textit{Gaussian disk, gap} model, the third row shows the \textit{Gaussian, disk, gap, ring} model, and the fourth and final row shows the \textit{Flat-Topped Gaussian} model. \label{fig:bestfit_profiles}}
 \end{figure*}

  \begin{figure*}[!ht]
    \hspace{30mm}\textbf{\underline{Gaussian disk residuals}} \hspace{45mm}\textbf{\underline{Gaussian disk, gap residuals}}
    \gridline{\fig{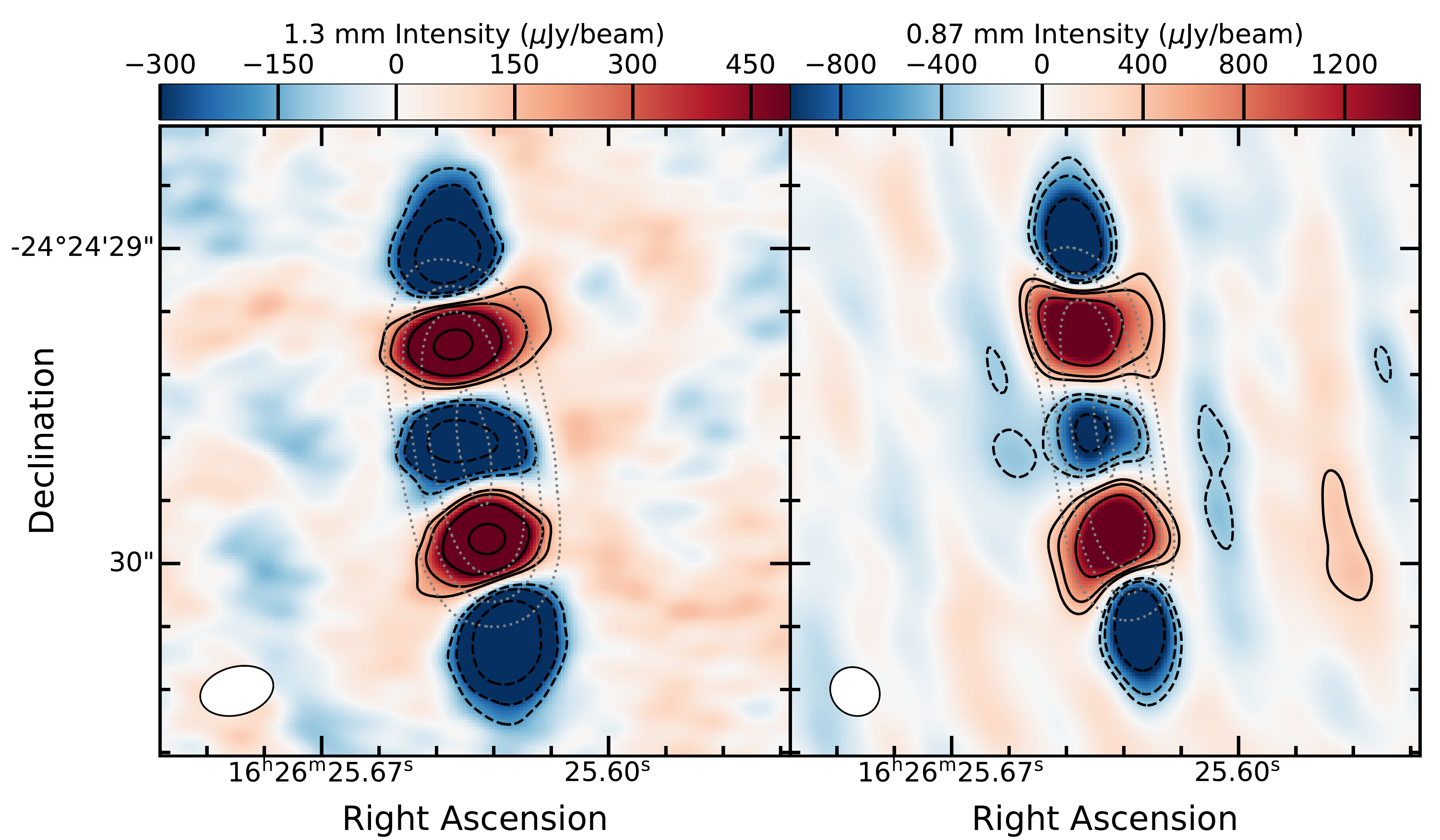}{0.5\textwidth}{}
          \fig{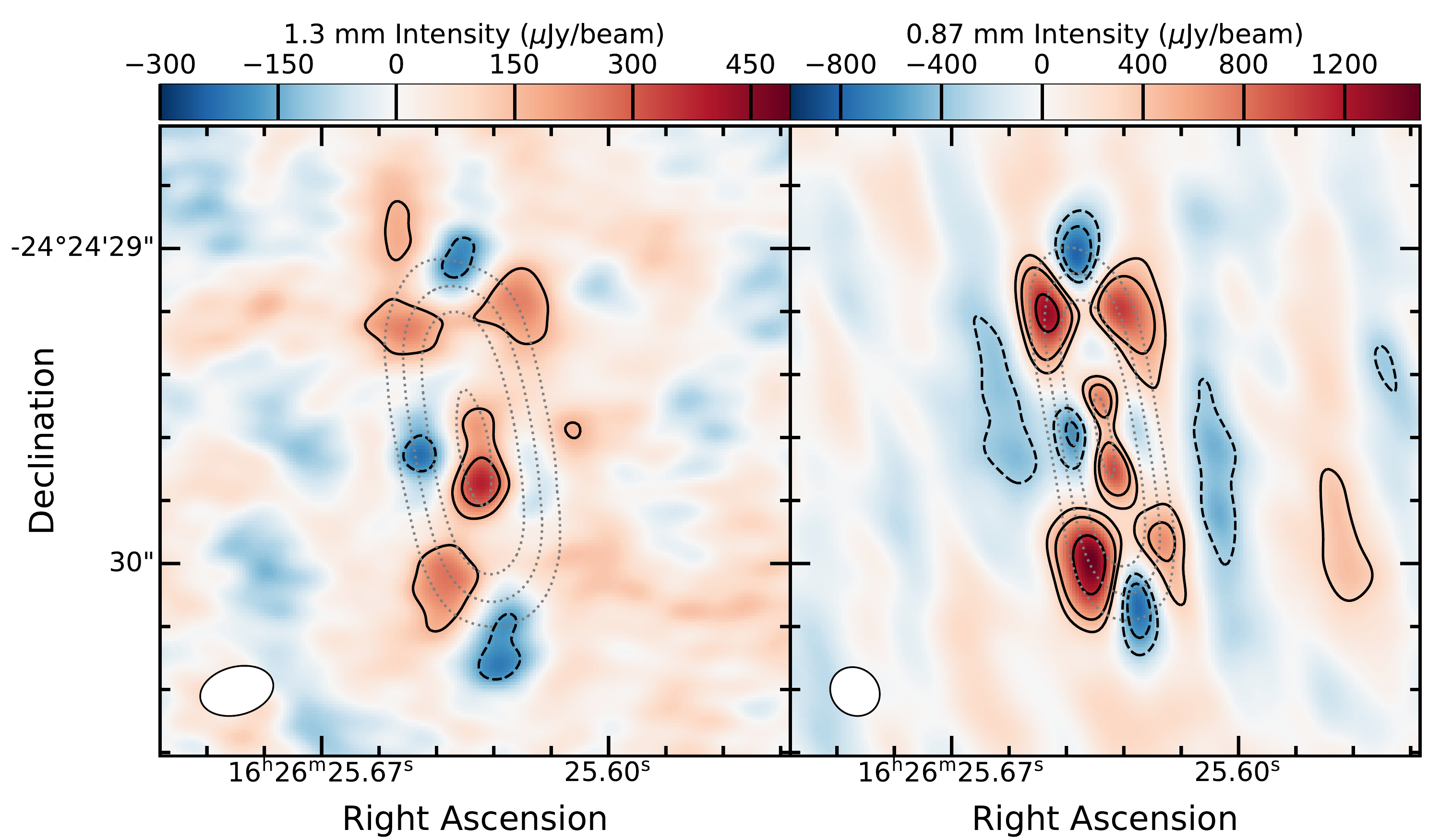}{0.5\textwidth}{}}
    \vspace{-4mm}
    \hspace{20mm}\textbf{\underline{Gaussian disk, gap, ring residuals}} \hspace{35mm}\textbf{\underline{Flat-Topped Gaussian residuals}}
    \gridline{\fig{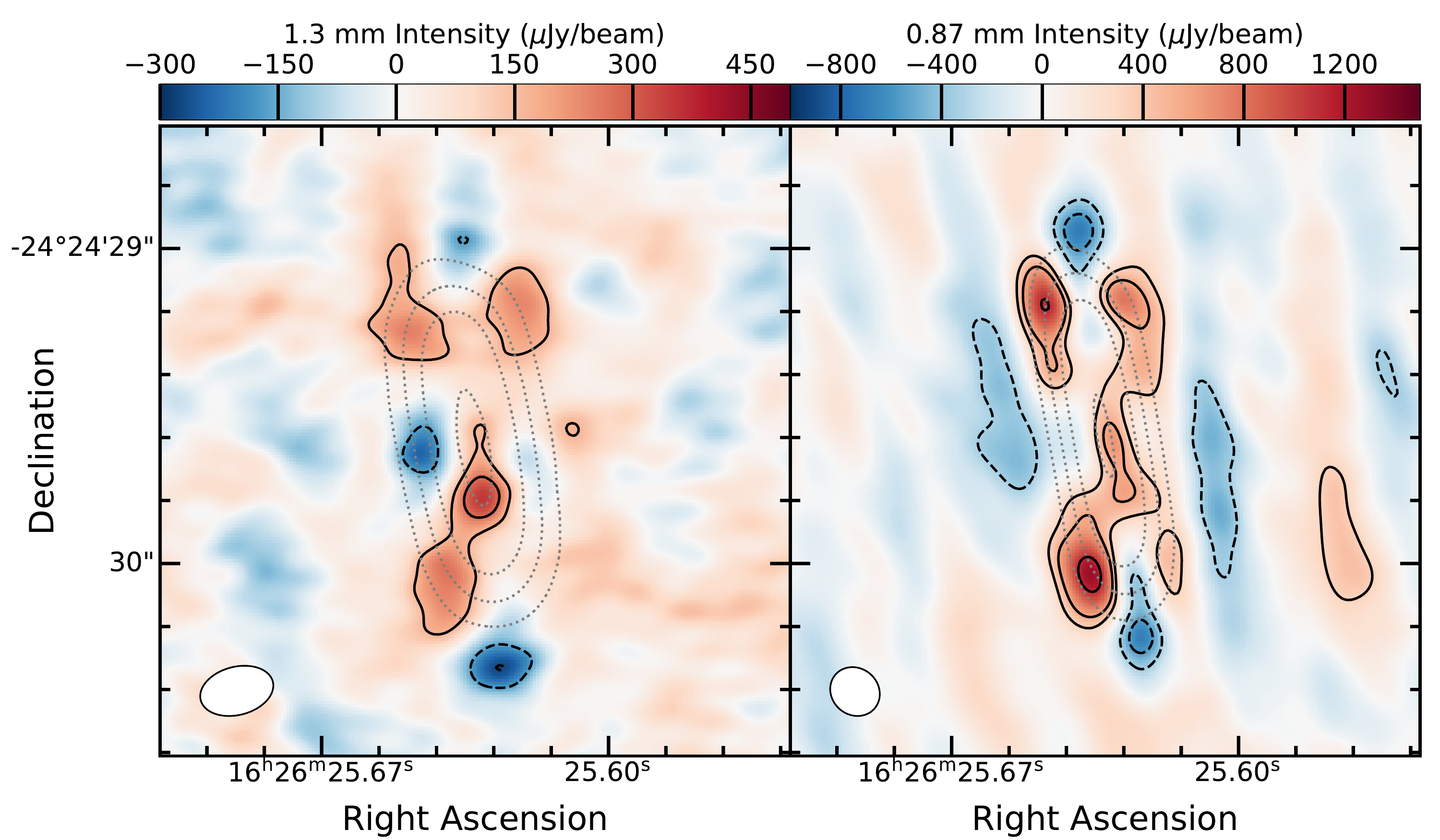}{0.5\textwidth}{}
          \fig{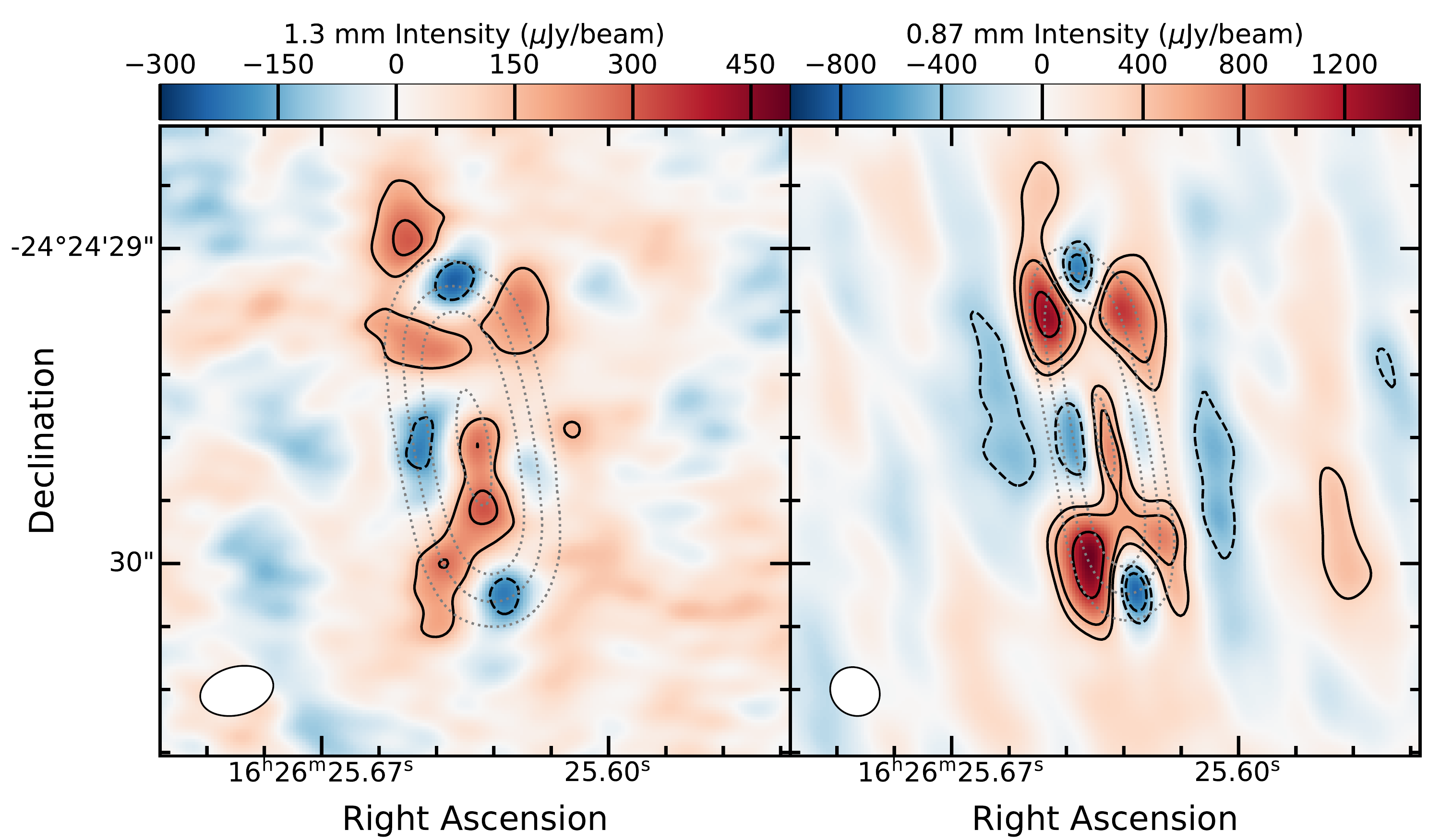}{0.5\textwidth}{}}
    \vspace{-5mm}
    \caption{Residuals were imaged after the subtraction of the Fourier transform of the different disk models. The black contours represent $\pm 3, 5, 10, 20\sigma$ residuals.The grey dotted lines represent the original disk emission at $20, 50, 100, 200\sigma$. In each pair of panels, the left panel represents the 1.3~mm residuals with a $\sigma_6 = 54\mu$Jy, and the right panel is the 0.87~mm residuals with a $\sigma_7 = 110\mu$Jy. Aside from the Gaussian disk model, the three other model residuals are remarkably consistent and similar, albeit with small differences, and generally display positive residuals at the disk's four corners. \label{fig:residuals}}
 \end{figure*}
 
Figure~\ref{fig:bestfit_profiles} shows the best-fit model \textit{uv} profiles compared to the 1.3 and 0.87~mm \textit{uv} data. The \textit{uv} data is shown in 55~k$\lambda$ bins at 1.3~mm and in 35~k$\lambda$ bins at 0.87~mm. Each row of panels displays a different geometric model fit to the data-sets and includes a zoomed-in section of the region where the Real \textit{uv} data are structured. The geometric models are plotted using the best-fit parameters found in Table~\ref{tbl:bestfits}, these are the median values of their posterior distributions with the uncertainties representing the 68\% inclusion interval. In general, the disk models unanimously find the same center point for the source RA = 16h 26m 25.6315s and Dec = $-24^{\circ}$ $24^{\prime}$ $29.6184^{\prime\prime}$ at 1.3~mm, RA = 16h 26m 25.6315s and Dec = $-24^{\circ}$ $24^{\prime}$ $29.5866^{\prime\prime}$ at 0.87~mm, position angle ($10.3\pm0.1^{\circ}$), and high inclination ($80.3\pm0.1^{\circ}$) confirming previous results \citep[e.g.,][]{Harris_2018,Sadavoy_2019}.

Figure~\ref{fig:residuals} shows the imaged residuals from each best-fit model using the same imaging procedure as described in Section~\ref{sec:almaobs}. The residuals are evaluated by subtracting the model's synthetic visibilities from the observations. Figure~\ref{fig:residuals} is thus useful to compare the impact each model has in the image plane, leaving residuals of varying distribution, morphology, and significance. Excluding the standard \textit{Gaussian disk} case case (upper left panel in Figure~\ref{fig:residuals}), the other three models show similar positive residuals at the four corners of the disk with the residuals more pronounced at the north and south corners of the eastern side of the disk compared to the western side. There are also negative residuals at the north and south ends along the disk's major axis. These residuals imply that there is excess emission off the disk mid-plane, such that the disk could be geometrically thick or flared (see Section~\ref{sec:toy}).

The standard \textit{Gaussian disk} model does not match the Real \textit{uv} visibilities. The residuals include both significant positive and negative artifacts, up to $23 \sigma_6$ and $23\sigma_7$. We thus focus on the other geometric models to address these features. The \textit{Gaussian disk, gap} and \textit{Gaussian disk, gap, ring} models fit the Real \textit{uv} data better and decrease the residuals to ${\sim}7\sigma_6$ and ${\sim}12\sigma_7$. The \textit{FTG} \textit{uv} visibilities and residuals are largely comparable to these structured disk models. In this case, the \textit{FTG} has a sharper disk edge that produces ringing similar to what is seen in the \textit{uv} visibilities. Since VLA 1623W has an ${\sim}80^{\circ}$ inclination, we are primarily observing the disk edge-on and as such, the dips and peaks in the Real visibilities may not be due to disk structures like gaps or rings. 

\subsection{Model quality}\label{sec:stats}
 
We use statistical tests including the reduced $\chi^2$, the Akaike Information Criterion \citep[AIC;][]{Akaike_1974}, and the Bayesian Information Criterion \citep[BIC;][]{Schwarz_1978}, to qualitatively compare the models described in Section~\ref{sec:model_fit_res}. The AIC and BIC both address the quality of the fit but also the profile complexity. They penalize, albeit differently, the increasing number of free parameters. The AIC is defined as $\text{AIC} = 2k + n\ln(RSS/n)$, where $k$ is the number of free parameters, $n$ the number of data points, and $RSS$ is the residual sum of squares. The BIC is evaluated as $\text{BIC} = k\ln(n) + n\ln(RSS/n)$. The inclusion of extra parameters inflicts a more severe penalization for the BIC compared to the AIC. Evaluating $\Delta$AIC and $\Delta$BIC between models (i.e. $\Delta$AIC = AIC$_0$ - AIC$_1$) provides a qualitative comparison. \citet{Kass_1995} quantifies $3 < \Delta\text{BIC} < 20$ to be positive evidence in favor of the model with the \emph{lower} BIC value, $20 < \Delta\text{BIC} < 150$ as strong evidence, and $>150$ as decisive evidence. We use the same comparison scale for $\Delta$AIC.

\begin{table}[!ht]
\begin{center}
\caption{Disk profiles}
\label{tbl:stats}
\begin{tabular}{l|lcc}
\hline
\hline
& \multicolumn{3}{c}{Best-fit Statistics} \\
Profile & $\lambda$ & 1.3~mm & 0.87~mm \\
\hline
Gaussian Disk & $\chi^2$ & 5.45 & 11.54 \\
& $\Delta$AIC & -1462 & -5463 \\
& $\Delta$BIC & -1452 & -5452 \\
\hline
Gaussian Disk& $\chi^2$ & 5.30 & 11.42 \\
Gap & $\Delta$AIC & -15 & -289 \\
& $\Delta$BIC & -33 & -312 \\
\hline
Gaussian Disk, & $\chi^2$ & 5.30 & 11.41 \\
Gap, & $\Delta$AIC & -20 & 96 \\
Ring & $\Delta$BIC & -64 & 40 \\
\hline
Flat-Topped & $\chi^2$ & 5.30 & 11.41 \\
Gaussian & AIC$_0$ & 87612 & 1208236 \\
& BIC$_0$ & 87665 & 1208315 \\
\hline
\end{tabular}
\end{center} 
\textit{Notes:} AIC$_0$ and BIC$_0$ are the reference values from the \textit{FTG} model used to compare with the other models. These are relative comparisons between pairs of models and can thus be applied to any pair to compare which model is superior.
\end{table}

\begin{figure}[!ht]
   \centering
   \vspace{-1mm}
    \hspace{10mm}\textbf{\underline{Intensity fits}}
    \vspace{-3mm}
    \gridline{\fig{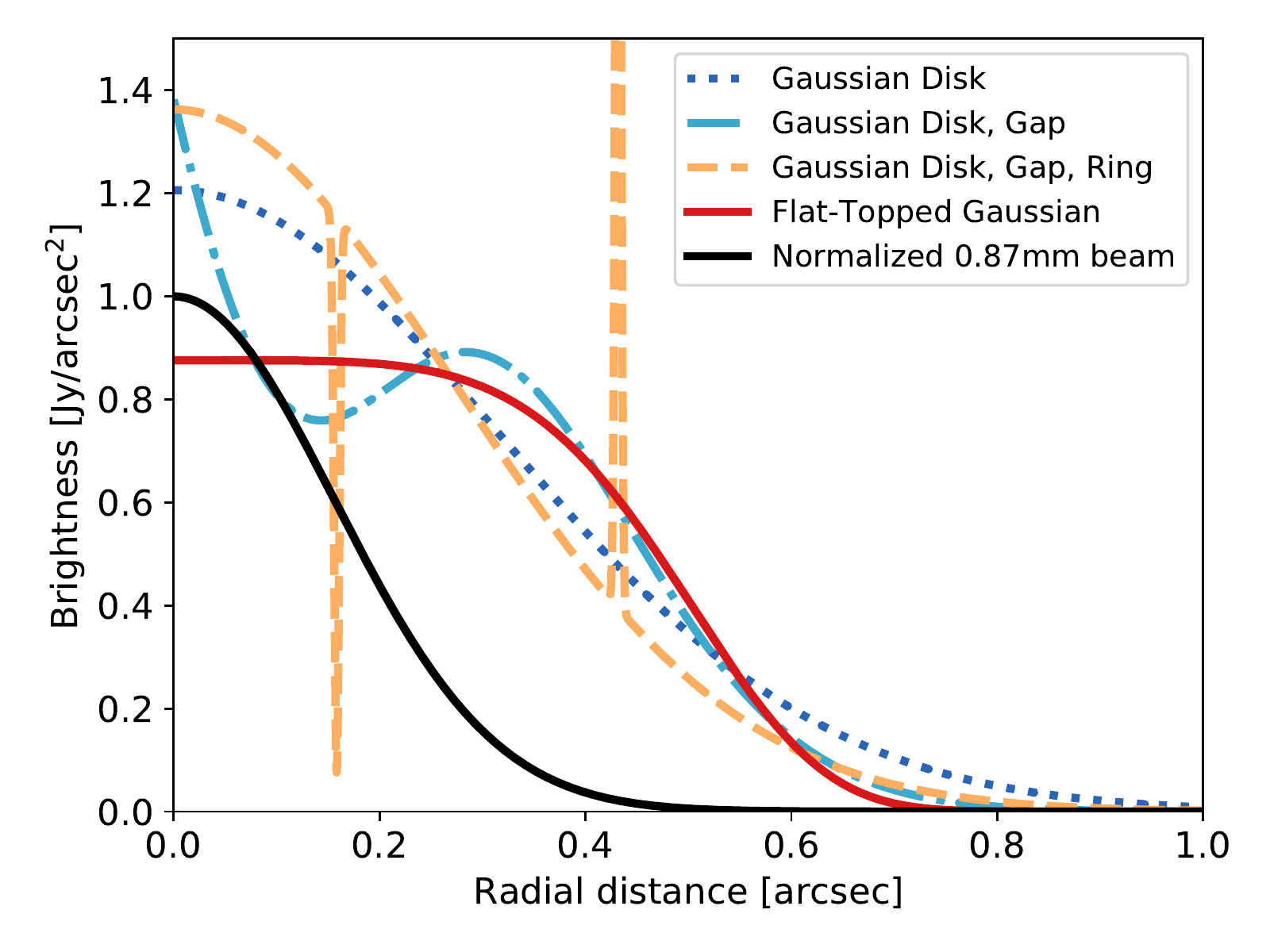}{0.46\textwidth}{}}
    \vspace{-5mm}
    \hspace{10mm}\textbf{\underline{\textit{uv} profiles}}
    \vspace{-3mm}
    \gridline{\fig{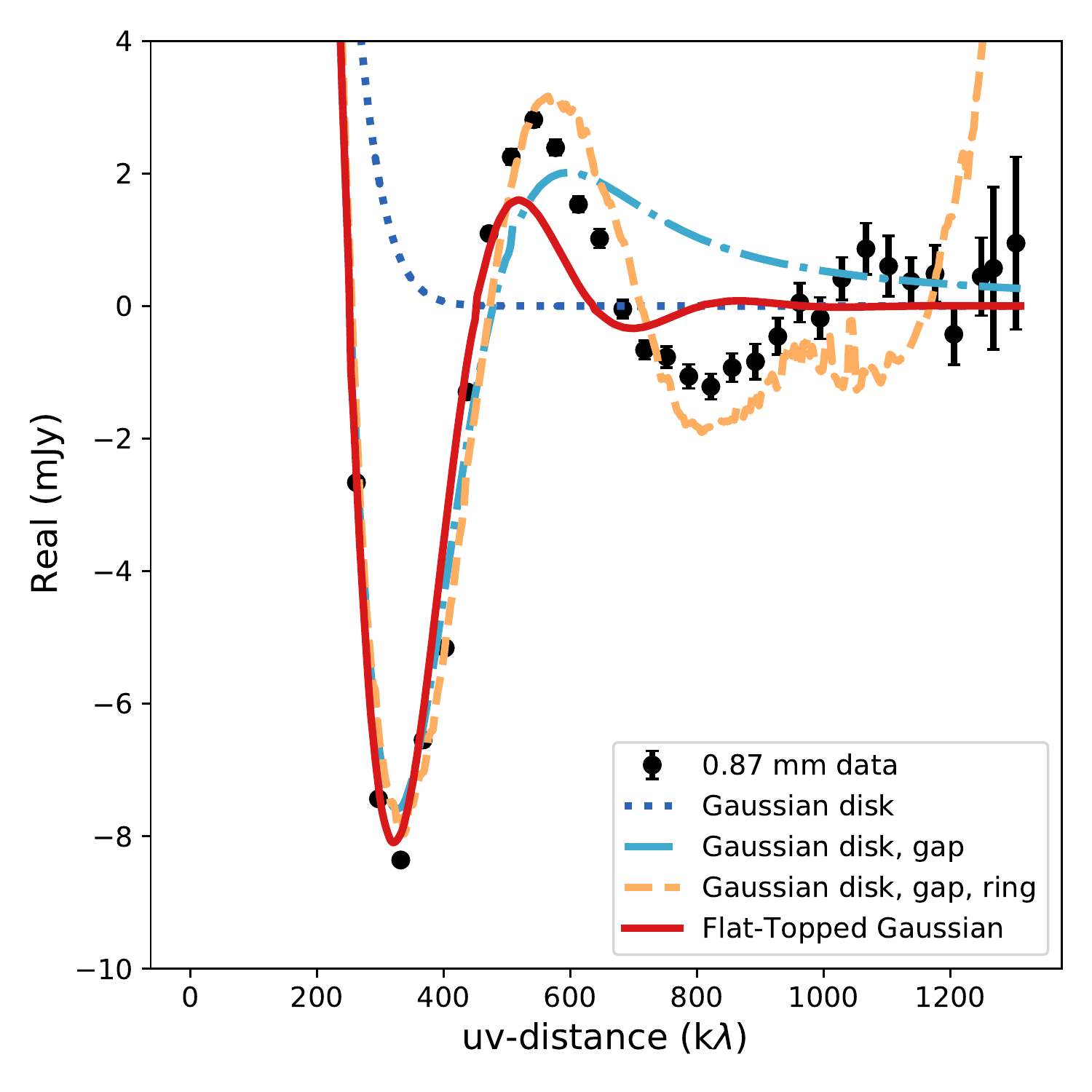}{0.46\textwidth}{}}
    \vspace{-8mm}
   \caption{ The variety of morphologies from the best-fit geometric models for the 0.87~mm data are shown here. \textbf{Top:} The intensities as a function of radial distance are shown. The solid black curve represents the normalized 0.87~mm circularized Gaussian beam thus illustrating the approximate image resolution. \textbf{Bottom:} The corresponding \textit{uv} profiles of the best-fit geometric models at 0.87~mm with the observations plotted in black. These are the same profiles shown as red lines in the right-most panels of Figure~\ref{fig:bestfit_profiles}.}
      \label{fig:modelsB7}
\end{figure}

Table~\ref{tbl:stats} shows the results from the statistical tests. We benchmark $\Delta$AIC and $\Delta$BIC against the \textit{Flat Topped Gaussian} where the reference AIC$_0$=87612 and BIC$_0$=87665 at 1.3~mm and AIC$_0$=1208236 and BIC$_0$=1208315 at 0.87~mm. Unsurprisingly, we find that the \textit{Gaussian disk} model does a significantly poor job fitting the observations based on the AIC and BIC parameters compared to the \textit{FTG}. The \textit{Gaussian disk, gap} model is also strongly disfavored based on both the AIC and BIC metrics. The \textit{Gaussian disk, gap, ring} model, however, cannot be fully ruled out as there is strong evidence in favor of it at 0.87~mm. 

Nevertheless, the \textit{Gaussian disk, gap, ring} model is not favoured at 1.3~mm and the resulting best-fit profile is unlikely to be observed with the resolution of the 0.87~mm observations. The best-fit ring and gap have widths of order ${\sim}0.3$ au (${\sim}2$ mas) which is 75 times lower than the beam resolution. Figure~\ref{fig:modelsB7} compares the best-fit geometric models studied here both plotted as brightness as a function of radius (top panel) and Real visibilities as a function of \textit{uv}-distance. These two panels highlight the extreme conditions for the \textit{Gaussian disk, gap, ring} model compared to the other, more smooth, profiles. Such sharp features would be surprising compared to the properties of gaps and rings seen in other Class 0/I protostellar disks \citep[][]{Segura-Cox_2020,Sheehan_2020}. We suggest that this fit deviates from observable solutions to purely addressing the features found in the Real \textit{uv} visibilities. Furthermore, the 1.3 and 0.87~mm structured disk fits are drastically different. In contrast, the \textit{Flat-Topped Gaussian} is statistically favored in nearly every comparison, the best-fits are consistent at both wavelengths, and it is also the simplest model that addresses the observed data as well as our expectation of a highly inclined optically thick protostellar disk. 

We compare the VLA 1623W observations and best-fit models in \textit{uv} visibility space since the beam-convolved images are very similar due to the large beam-size relative to the typical disk structure sizes identified by the best-fit models in Table~\ref{tbl:bestfits} (see Appendix~\ref{sec:appendix_imagespace} for more details). Since the \textit{uv} visibilities are not convolved with the beam, the differences in the models are detectable and we can use statistics tests to distinguish between them.

Regardless of the model, four symmetric positive residual features are generally found at the edges of the disk consistently across both wavelengths, see Figure~\ref{fig:residuals}. These residuals vary in their significance depending on the model that is subtracted from the observations but remain similar in their morphology and distribution. These residuals suggest there are features of the disk that are not captured by our simple geometric models and assumptions.

\subsection{Toy model}\label{sec:toy}

The symmetry, location, and consistency of the residuals across the disk models and both wavelengths suggest a real feature of the disk. Given the high inclination of VLA 1623W (${\sim}80^{\circ}$), the residuals may be evidence of the outer disk vertical scale height. For young disks the outer edge may still be puffed up \citep[e.g., IRAS 04302;][]{Villenave_2020}, because the millimeter-sized dust has yet to settle. The large dust is expected to settle eventually onto the thin, cold mid-plane as observed in more evolved edge-on Class II protoplanetary disks \citep{Villenave_2020,Villenave_2022}.

\begin{figure}[!ht]
   \centering
   \vspace{-1mm}
    \hspace{10mm}\textbf{\underline{Edge-on flared disk toy model}}
    \gridline{\fig{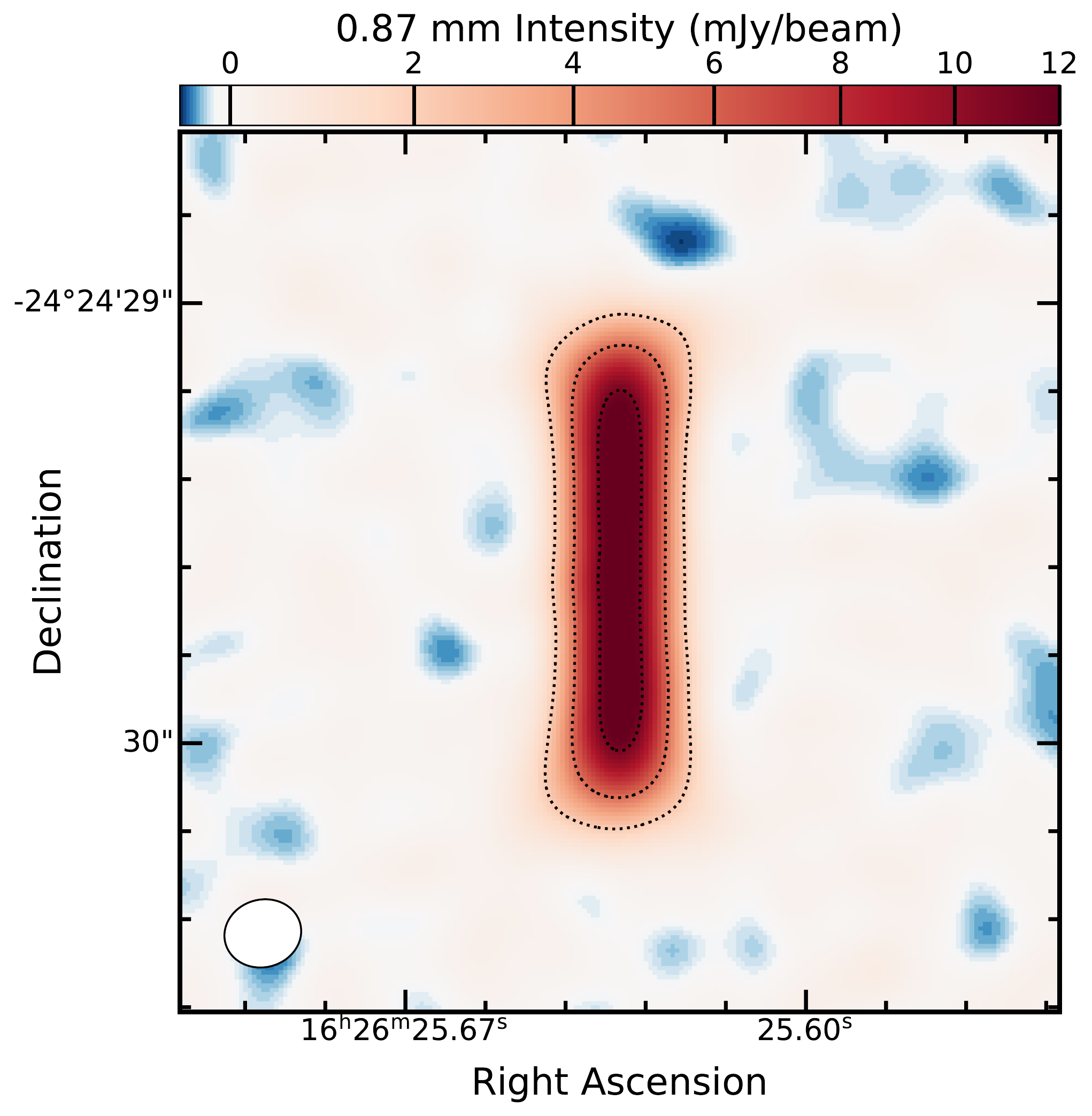}{0.43\textwidth}{}}
    \vspace{-5mm}
    \hspace{10mm}\textbf{\underline{Residuals}}
    \gridline{\fig{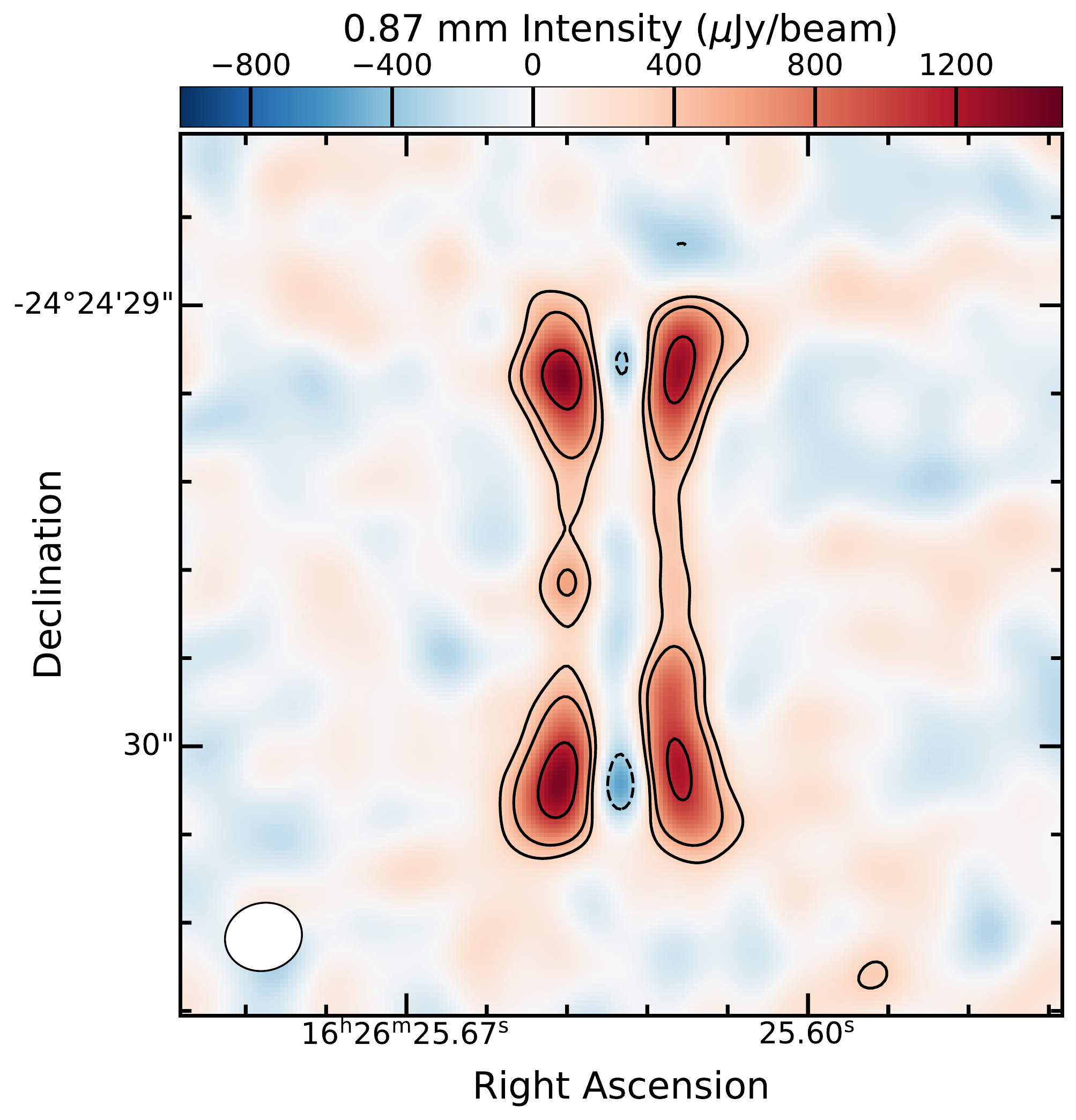}{0.43\textwidth}{}}
    \vspace{-8mm}
    \caption{\textbf{Top:} Simulated observations at 0.87~mm of the flared edge-on disk toy model image using the same observational conditions as VLA 1623W. The parameters used are $h_{\text{off}}=4$ au, $h_0=0.1$, $\beta=5$, and $R_0=30$ au. The black dotted lines represent the contours of the 0.87~mm model disk emission at 20, 50, 100 $\sigma_7$. \textbf{Bottom:} Imaged residuals after the subtraction of the Fourier transform of the best-fit \textit{Flat-Topped Gaussian} model from the simulated observations of the toy model (Top). These imaged residuals correspond to the central panel in Figure~\ref{fig:grid_4AU} as it is the qualitatively most similar to the observation's residuals. The black contours represent $\pm 3, 5, 10, 20\sigma_7$ residuals.}
     \label{fig:toymodel}
\end{figure}

Since \texttt{GALARIO} is limited to 2D modeling \citep{Tazzari_2018}, we build a simple completely edge-on ($90^{\circ}$) flared disk toy models based on the best-fit \textit{FTG} model at 0.87~mm. Briefly, we assume that the disk scale height (seen as a width when edge-on) increases as a function of radius. The key parameters used to describe the toy model are encapsulated in the vertical disk scale height parametrization, that is $\sigma_h$ of a Gaussian function such that
 \begin{equation}
     \sigma_h(R) = h_0 \left(\frac{R}{R_0}\right)^{\beta} + h_{\text{off}}
     \label{eq:sigmah}
 \end{equation}
\noindent where $h_0$ is the height scaling factor, $R_0$ is the radial scaling factor dictating at which radii flaring becomes important, $\beta$ is the flaring power, and $h_{\text{off}}$ is added to account for both the geometric thickness of the disk \citep[e.g., it is not necessarily razor-thin like Oph163131;][]{Villenave_2022} as well as thickness in the disk from not being truly at $90^{\circ}$ inclination. We inject random Gaussian noise to closely match the sensitivity of the observations. Further details about the toy model's description can be found in Appendix~\ref{sec:appendix_toygrid}. 

We take a grid of toy models and fit them with an edge-on \textit{FTG} model using \texttt{GALARIO} to simulate our observations and then image the residuals of the toy model minus the best-fit edge-on \textit{FTG} model. The residuals in this case should represent the remaining emission that is offset from the disk mid-plane that is not capture by a flat edge-on \textit{FTG} model. Therefore, a good toy model should reproduce the observed residual features seen in Figure~\ref{fig:residuals}. Figure~\ref{fig:toymodel} shows a sample disk model (top) and the corresponding imaged residuals (bottom), see Appendix~\ref{sec:appendix_toygrid} for additional tests. 

With the {edge-on flared protostellar disk} toy model, we can reproduce a similar residual morphology at the four corners of the disk. However, we emphasize we are not fitting the observations, rather we are making a qualitative comparison. While the residuals show qualitatively similar positive and negative residuals along the north and south, we subtract slightly too much emission along the major axis and not enough emission along the minor axis. Since the toy model assumes a perfectly edge-on \textit{FTG} disk input whereas VLA 1623W's disk inclination is ${\sim}80^{\circ}$, some of these differences could be the effect of inclination. In addition, the toy model is a simple approximation of the dust emission, whereas a full description would require radiative transfer modeling and directly fitting for the disk structure simultaneously, especially as we would anticipate a change in optical depth and temperature from the mid-plane to the flared edges and there may be non-negligible contributions to the intensity from scattered emission. Such 3D radiative transfer models are beyond the scope of this work but will be necessary to fully evaluate the features seen in VLA 1623W.

\section{Discussion}\label{sec:discussion}

From our model fits and toy model, we propose that VLA 1623W is indeed a protostellar disk and that the disk is both optically thick and flared. The lack of a large envelope feature \citep{MurilloLai_2013,Kirk_2017}, suggests that the millimeter emission purely traces the disk.  Within the VLA 1623 protostellar system, both the VLA 1623Aa \& Ab canonical protobinary and VLA 1623B are considered Class 0 sources. VLA 1623W, however, has been labeled as a Class I source based on its spectral energy distribution and its envelope-to-star mass ratio \citep{MurilloLai_2013,Murillo_2018}. \citet{Harris_2018} find that West could have been ejected from the triple system VLA 1623Aa, Ab, \& B based on proper motion analysis, thus possibly implying a common age. In an ejection scenario, the envelope could be stripped such that only the disk survives \citep{Reipurth_2000}. The failure of the models that include an envelope component (e.g., the \textit{PLCT} model; see Section~\ref{sec:models}) suggests there is no significant envelope in these deep mm-observations either. We acknowledge that the envelope component could be encapsulated at \textit{uv} distances $<50$~k$\lambda$. To investigate this further, we fit the \textit{PLCT} model to data-sets where VLA 1623 Aa, Ab, \& B have been subtracted but still contain the full \textit{uv} range. The same result is found as with the 50~k$\lambda$ cut data where the \textit{PLCT} model turns to a Gaussian disk model with no envelope being found at both 1.3 and 0.87~mm. Thus, VLA 1623W could be as young as the VLA1623 Aa, Ab, \& B sources but simply lacks an envelope due to being dynamically ejected. Ultimately, VLA 1623W's exact classification is not clear, but it is likely of similar age to the VLA 1623Aa, Ab, \& B system.

\subsection{Vertical settling}\label{sec:settling}

The disk flaring observed in millimeter dust for VLA1623W implies that the disk's outer edges are still significantly puffed up with vertically distributed mm-dust above the cold disk mid-plane. This dust has not yet completely settled onto a thin disk. In general, vertical settling is observed in the difference in scale heights between the infrared and millimeter-sized dust in protoplanetary disk observations \citep[e.g.,][]{Villenave_2020, Miotello_2022}. The small micron-sized dust follows the higher gas scale height whereas the millimeter-sized dust appears to be well settled in the mid-plane \citep{Barriere_2005}. 

Vertical settling into a very thin dust disk has been observed for some Class II protoplanetary disks \citep[e.g., HL Tau, 2MASS J16083070-3828268, Tau 042021, SSTC2DJ163131.2-242627, Oph 163131;][]{Pinte_2016, Villenave_2019, Villenave_2020, Villenave_2022, Wolff_2021}. By contrast, some disks and/or parts of disk substructures have been found to show elevated millimeter dust vertical scale heights. For example, HD 163296, a Class II disk where \citet{Doi_2021} find both settled and puffy dust rings, IRAS 04302 a Class I protostellar disk where \citet{Villenave_2020} suggest that it may be flared and less settled, and the Class 0 protostellar source HH 212 which exhibits hamburger-shape emission indicative of the unsettled millimeter-sized dust \citep{Lee_2017}. Both \citet{Ohashi_2022_L1527} and \citet{Sheehan_2022_L1527} report flaring for the edge-on Class 0 disk L1527 IRS in millimeter observations, implying that large dust grains are elevated above the disk mid-plane. These observations seem to indicate that some protostellar disk may be puffed up with longer dust settling timescales. For example, significant turbulence may stir up large dust \citep{Sheehan_2022_L1527}. 

\begin{figure*}[!ht]
    \centering
    \gridline{\fig{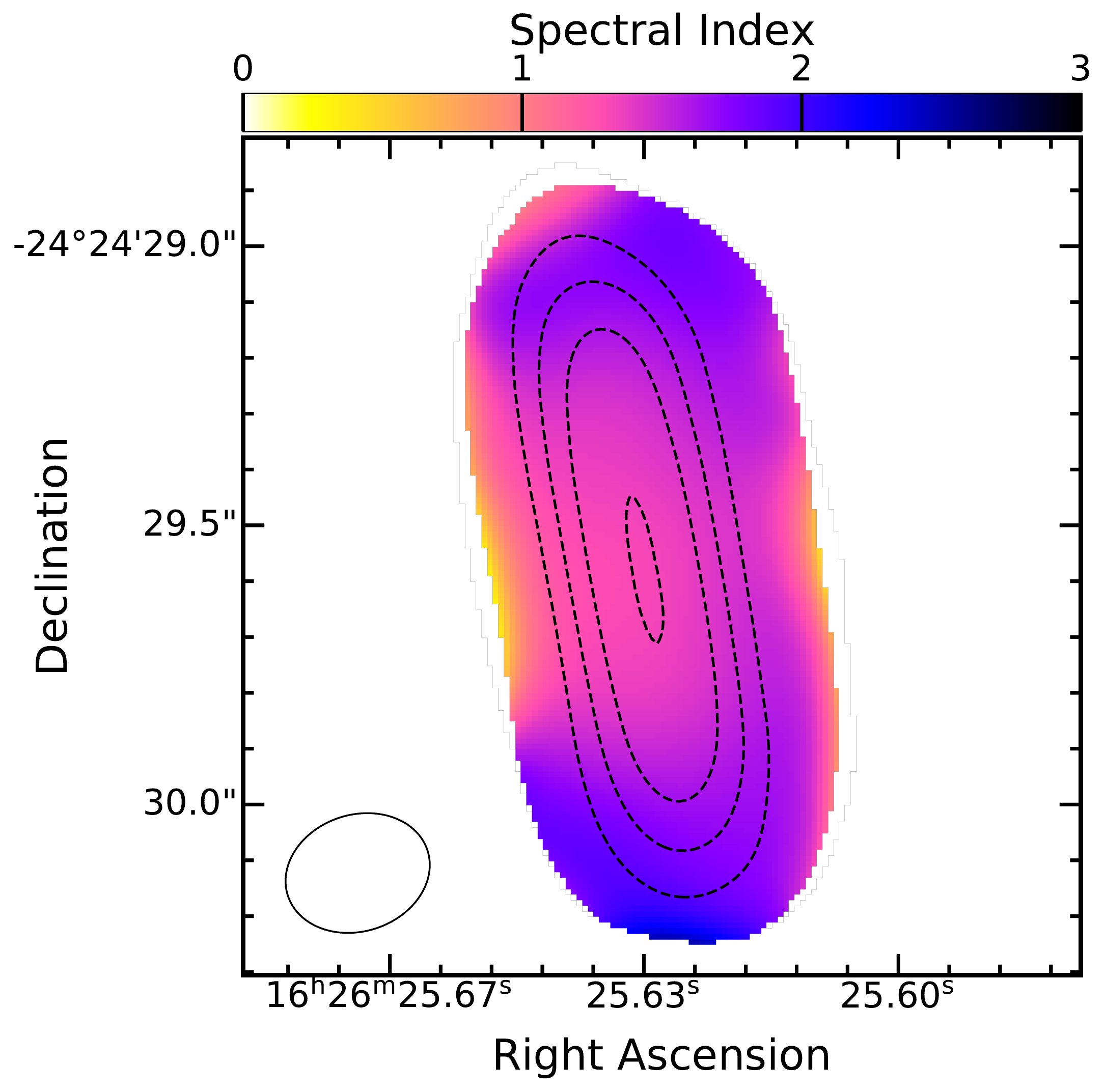}{0.49\textwidth}{}
          \fig{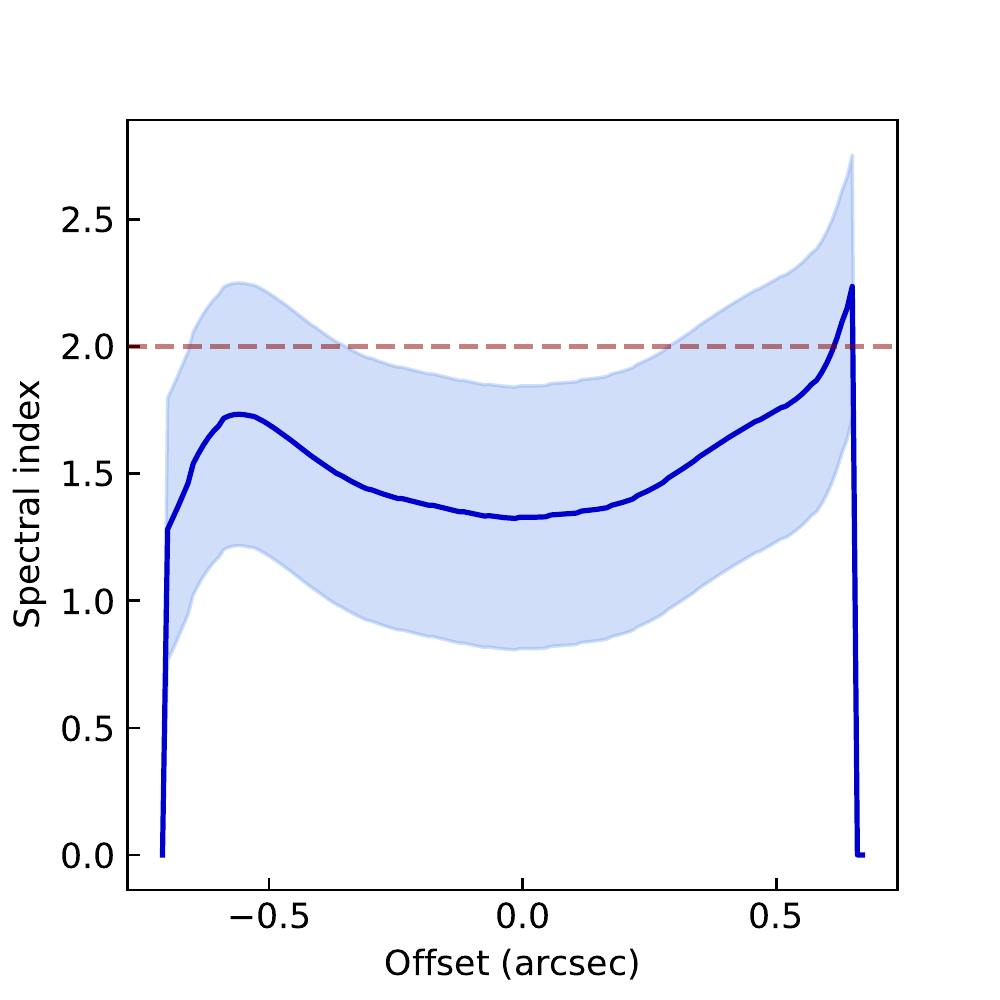}{0.49\textwidth}{}
          }
      \caption{\textbf{Left:} Map of spectral index for the region where the intensity is greater than 10$\sigma$ both at Bands 6 and 7. The dotted lines represent the contours of 0.87~mm disk emission of VLA 1623W at 20, 50, 100, 200$\sigma_7$ smoothed to the same resolution as the spectral index map. \textbf{Right:} Spectral index along the disk's major axis. The x-axis is the offset from the disk center, from the Flat-Topped Gaussian fits, we find $\sigma_D\sim0.45-0.46^{\prime\prime}$. Shaded light blue is the spectral index evaluated assuming a 10\% flux uncertainty from calibration for both 1.3 and 0.87~mm data.}
         \label{fig:spindex}
 \end{figure*}
 
The current methodology used provides a good first look into VLA 1623W's morphology and its features. We highlight the uniqueness of finding possible flaring in a Class 0/I disk, adding to the small sample of protostellar disks known to exhibit similar behavior \citep[HH 212, IRAS 04302;][]{Lee_2017,Villenave_2020}. Furthermore, the high inclination challenges the analytic profiles which do not account for scattering and optical depth \citep{Tazzari_2018}. In Figure~\ref{fig:residuals}, we may be seeing some of these effects as the residuals in the north and south corners of the eastern side of the disk are consistently brighter by a factor of ${\sim}2$ compared to the equivalent locations on the western side.

This difference in brightness could imply that the eastern side is slightly warmer than the western side. For example, if the eastern side is the far side of the disk, we may be observing the passively heated dust in the flared disk whereas for the western side (near side) the disk may be shadowed by the cooler outer edge of the flare \citep[e.g., similar to L1527 IRS][]{Ohashi_2022_L1527}. Alternatively, VLA 1623W has detected polarized light of order 1-1.5\% \citep{Harris_2018,Sadavoy_2019}. If the polarization is from dust self-scattering, there will be a flux enhancement in the direction of forward scattering over backward scattering \citep[e.g.,][]{Perrin_2015,Yang_2017}. In this scenario, the brighter eastern side would be the near-side of the disk. Thus, temperature and scattered light present competing effects that may both be present. 3D radiative transfer calculations with a flared disk model will be necessary to discern the reason for the east-west residual asymmetry.

The toy model is not designed to reproduce the observations. We emphasize that we are not aiming to quantitatively fit the toy model to the data but rather we qualitatively compare the residuals using an identical procedure. From the results (see Figures~\ref{fig:residuals} and~\ref{fig:toymodel}), the residual's morphological similarities are striking and thus appear as a possible solution to VLA 1623W disk's features. 

Further work, constraining the rate at which the millimeter-sized dust concentrates in the mid-plane, is important to the description of the initial conditions for planet formation given that dust densities are a crucial component to trigger this process \citep{Drazkowska_2022,Miotello_2022}. Observing a larger sample of edge-on protostellar disks to connect with \citet{Villenave_2020} protoplanetary disks would allow for a more accurate evolutionary picture if the disk kinematics and envelope components can be properly constrained. 

\subsection{Spectral Index Map}\label{sec:spindex}

From the polarization results, we worked with the assumption that VLA 1623W is optically thick for the disk modeling. Now, with the two wavelengths available, we can test this assumption independently. The ratio of the millimeter continuum observations at different wavelengths provides an estimation of the spectral index $\alpha_{mm}$ as $S_{\nu} \propto \nu^{\alpha_{mm}}$. Across disks, $\alpha_{mm}$ has been used to constrain the dust optical depth and consequently describe the dust grain sizes, however these are degenerate \citep{Williams_2011, Miotello_2022}. The dust opacity is expected to have a power-law dependency at millimeter frequencies which corresponds to the Rayleigh-Jeans regime such that $\kappa_{\nu} \propto \nu^{\beta}$ where $\kappa_{\nu}$ is the dust absorption coefficient and $\alpha_{mm} = \beta + 2$ at the Rayleigh-Jeans limit (where $h_{\nu} / kT << 1$) such that optically thick dust has $\alpha_{mm}\rightarrow2$ or large grains that have $\beta=0$.

We measured the $\alpha_{mm}$ map using the Stokes $I$ continuum observations at 1.3 and 0.87~mm. We center the disks according to the best-fit \textit{FTG} model offsets obtained at each wavelength. We clean the observations with \textit{uniform} weighting across the same \textit{uv}-range (17-1700~k$\lambda$) and smoothed each map by the beam of the other so that the two images have a common resolution of $0.26\times0.21^{\prime\prime}$. To focus on the main disk emission, we mask out emission $<10\sigma$ at both wavelengths. Using the \texttt{immath} task in \texttt{CASA}, we evaluate the spectral index map pixel-by-pixel from the two wavelength observations.

Figure~\ref{fig:spindex} (left) shows the $\alpha_{mm}$ map with dashed contours of the VLA 1623W disk emission at 0.87~mm and (right) the radial profile of the $\alpha_{mm}$ along the disk's major axis. We find a global spectral index of $1.5\pm0.2$ consistent with a high dust optical depth across the disk and possibly lower than edge-on protoplanetary disks, e.g., a median $\alpha_{mm}$ of $2.5\pm0.3$ from  \citet{Villenave_2020}. The value of $\alpha_{mm}$ increases radially along the disk's major axis similar to \citet{Villenave_2020}. Generally, we find that VLA 1623W is consistent with being optically thick throughout most of the disk. 

The optically thick dust result from polarization and $\alpha_{mm}$ are consistent with the \textit{Flat-Topped Gaussian} description which implies a constant surface brightness across the disk, rather than a peaked center. We are observing the outer region of the disk more uniformly. This is particularly visible along the disk's major axis, which we are probing in better detail than the minor axis, due to the high inclination along the line of sight. Modeling the disk according to this prescription implies that we are primarily tracing the outermost dust surface layer of an optically thick source. Note that some protoplanetary disks, \citet{Villenave_2020} report similar sharp edges in the brightness profiles along the disk's major axis.

It must be noted that a further nuance exists with such inclined protostellar disks, that is the effect of scattering-induced intensity reduction along the disk mid-plane and possible enhancements in the puffy outer edges \citep[see review in][]{Miotello_2022}, which given the high inclination of VLA 1623W can impact the derived millimeter spectral index by decreasing the intensity in the optically thick region further \citep{Zhu_2019,Sierra_2020}. Furthermore, low protostellar disk temperatures in the outer layers due to self-obscuration will also contribute to decreasing the spectral index \citep{Zamponi_2021}.

\subsection{No disk substructure}\label{sec:structure}

We favour the \textit{Flat-Topped Gaussian} model over the more complex \textit{Gaussian disk, gap, ring} model even though the latter produced a statistically superior fit to the observations at 0.87~mm. As discussed in Section~\ref{sec:toy}, the best-fit parameters for the \textit{Gaussian disk, gap, ring} profile required a sharp gap and sharp ring, that are well below the current observation's resolution (see Figure~\ref{fig:modelsB7}). These sharp features artificially introduce ringing into the \textit{uv} visibilities that can mostly fit the observations but may instead be artifacts rather than real disk substructures.

These results indicate the dangers of over-interpreting structure in young and inclined protostellar disks. As structured disks have been readily observed at the Class II stage \citep[e.g.][]{ALMA_2015,Andrews_2016,Isella_2016,Andrews_2018,Huang_2018,Long_2018,vanderMarel_2021_asymmetries}, there is the need to identify the precursors to such features in the younger disks, namely in the protostellar Class 0/I stages. Some young large massive protostellar disks have indeed been observed to have substructure including IRS 63 in Ophiuchus \citep{Segura-Cox_2020}, L1489 IRS \citep{Ohashi_2022_L1489}, and a series of Class 0/I disks in Orion \citep{Sheehan_2020}. As the community searches for the precursors of the highly structured disks found in the DSHARP sample for example \citep{Andrews_2018,Huang_2018}, we must be wary of the evidence and models presented. The models presented here represent simple 1-D geometric profiles that cannot fit complex disks with flared structures well. As a further test of the disk models, we also use a non-parametric approach by fitting the data-sets with \texttt{frank} \citep{Jennings_2020}. The results from \texttt{frank} suggest a highly structured protostellar disk, see Appendix~\ref{sec:appendix_frank}, that is most similar to the \textit{Gaussian disk, gap, ring} model in Section~\ref{sec:model_fit_res}. However, it is difficult to quantitatively and statistically compare the model fits across both codes. Given, that we reject \texttt{GALARIO}'s complex \textit{Gaussian, disk, gap, ring} model, we are inclined to also disfavor \texttt{frank}'s highly structured disk. Both codes are limited to geometrically thin disks, whereas VLA 1623W appears more complex. Ultimately, 3D radiative transfer modeling will be necessary to capture the structure, vertical extent, and optical properties of this source.

\section{Conclusions}\label{sec:conclusion}

We fit VLA 1623W 1.3 and 0.87 millimeter observations using simple geometric models to determine if this source is consistent with standard disk models. Our main results and interpretations are as follows:

\begin{enumerate}[noitemsep]
    \item VLA 1623W is consistent with being a young protostellar disk. It is highly inclined, optically thick, and well-characterized by a modified flat-topped Gaussian, indicating that the emission is relatively constant along the major axis and not peaked.
    \item We find similar positive residuals at the four corners of the disk for various models and at both wavelengths. This morphology is well matched by a toy model of an edge-on flared disk. Thus, the large dust grains in VLA 1623W may not have had time to settle onto a thin midplane.
    \item For protostellar sources, structured Real \textit{uv} visibilities do not necessarily imply substructure. 
\end{enumerate}

This study takes advantage of the high sensitivity Stokes $I$ continuum data from polarization observations of disks \citep{Harris_2018,Sadavoy_2019}. The ability to discern subtle hints in the disk morphology such as the flaring and rule out the presence of substructure is instrumental in our descriptions of protostellar disks. The protostellar disk polarization observations thus support this goal in addition to constraining the dust grain sizes \citep{Kataoka_2015}.

If VLA 1623W is confirmed to be flared from future millimeter wavelengths at high-resolution and sensitivities, then it will provide an interesting laboratory to study dust settling as well as dust grain properties across the vertical disk scale height in protostellar sources, and could add to young puffed-up disks at millimeter wavelengths \citep[e.g., HH 212, l1527;][]{Lee_2017,Ohashi_2022_L1527,Sheehan_2022_L1527}. In turn, this would provide constraints on the dust settling rates from protostellar to protoplanetary disks, setting the stage for comparisons with highly settled Class II disks \citep{Villenave_2022}. As we continue to search for the initial precursors to disk substructure and aim to constrain planet formation's initial conditions, we must further enlarge our sample of protostellar disks at high-resolution and sensitivity. 
\vspace{3mm}

\textit{Software:} \texttt{CASA} \citep{McMullin_2007}, \texttt{GALARIO} \citep{Tazzari_2018}, \texttt{emcee} \citep{Foreman-MacKey_2013}, \texttt{matplotlib} \citep{Hunter_2007}, \texttt{corner} \citep{Foreman-MacKey_2016}, \texttt{APLpy} \citep{Robitaille_2012}.

\textit{Acknowledgments:} We thank the referees for their constructive and useful suggestions. The authors are grateful to Luca Matrà and Marco Tazzari for helpful discussions in running \texttt{GALARIO}. AM and SIS acknowledge support from the Natural Science and Engineering Research Council of Canada (NSERC), RGPIN-2020-03981. LWL acknowledges support from NSF AST-1910364 and NSF AST-2108794. ALMA is a partnership of ESO (representing its member states), NSF (USA) and NINS (Japan), together with NRC (Canada) and NSC and ASIAA (Taiwan) and KASI (Republic of Korea), in cooperation with the Republic of Chile. The Joint ALMA Observatory is operated by ESO,AUI/ NRAO and NAOJ. This paper makes use of the following ALMA data: 2015.1.01112.S and 2015.1.00084.S.

\bibliography{references}{}
\bibliographystyle{aasjournal}

\section{Appendix}\label{sec:appendix}

 \subsection{Best-fit geometric disk model beam-convolved images} \label{sec:appendix_imagespace}
 
\begin{figure*}[!ht]
 \centering
    \gridline{\hspace{12mm}\fig{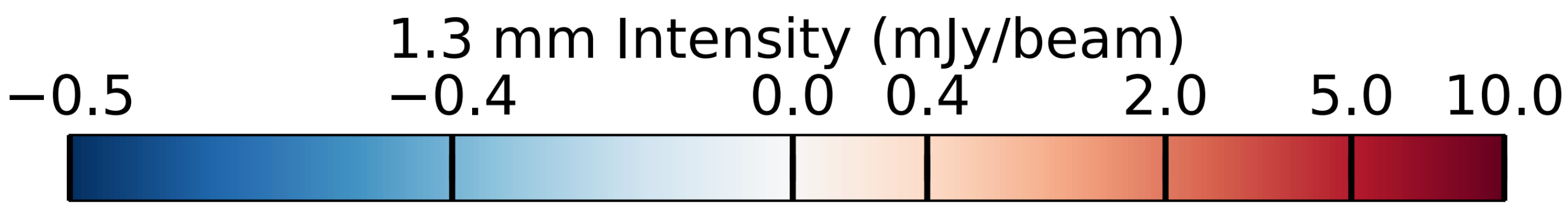}{0.45\textwidth}{}}
    \vspace{-10mm} 
    \gridline{\fig{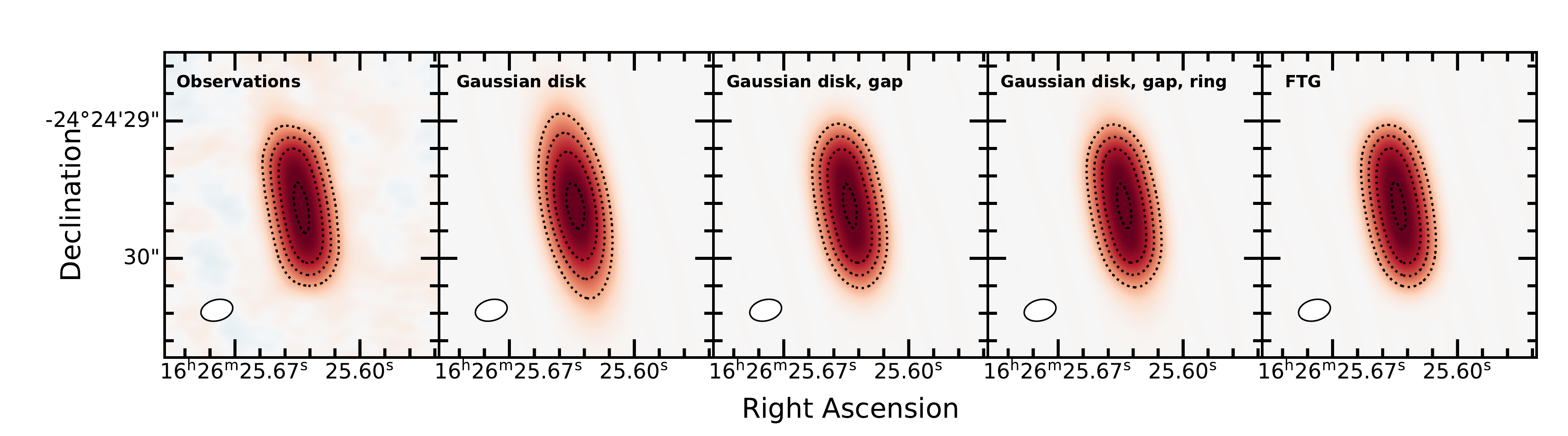}{0.99\textwidth}{}}
    \vspace{-5mm} 
    \gridline{\hspace{12mm}\fig{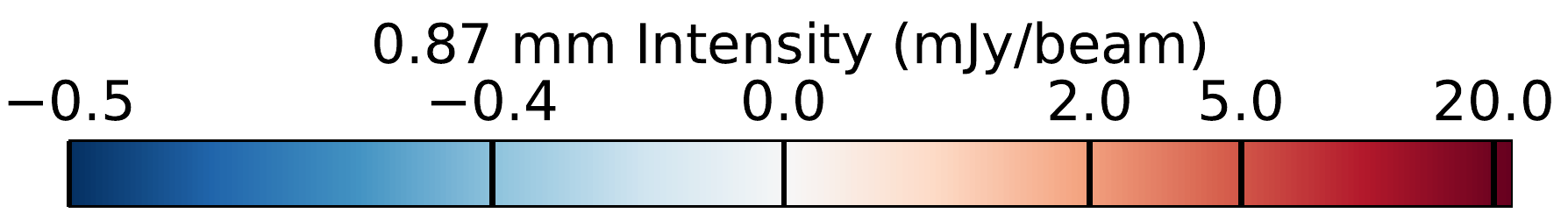}{0.45\textwidth}{}}
    \vspace{-9.5mm} 
    \gridline{\fig{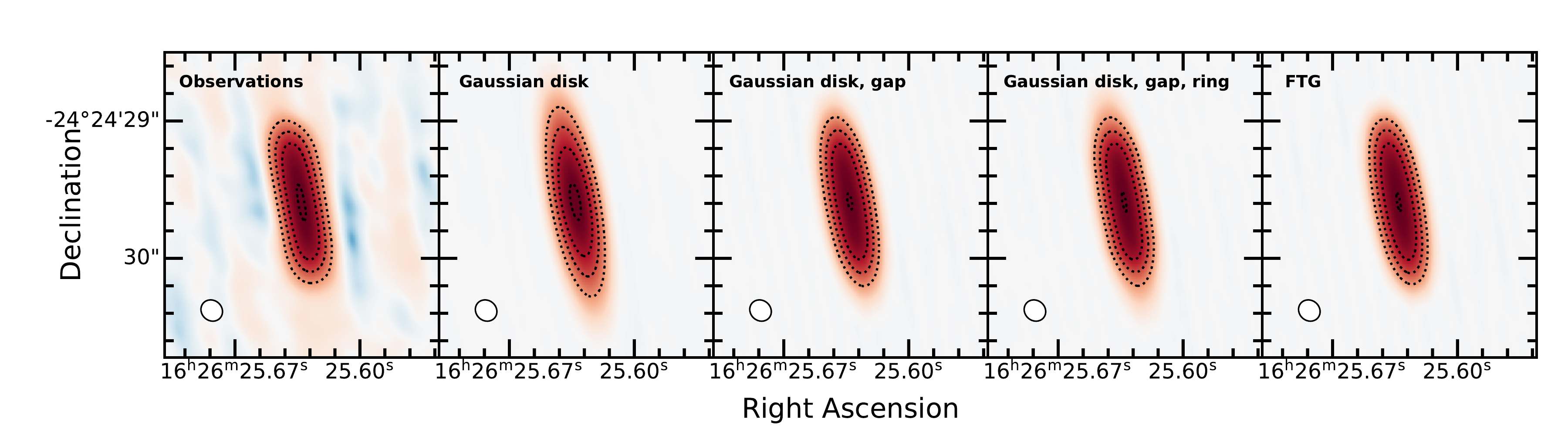}{0.99\textwidth}{}}
    \caption{Images of the best-fit disk models compared to the observations (leftmost panel). The model images were obtained by taking the Fourier transform of the best-fit disk model image and sampling the \textit{uv} plane of the observations using \texttt{GALARIO}. \textbf{Top:} At 1.3~mm and \textbf{Bottom:} at 0.87~mm.}
    \label{fig:convolved_images}
 \end{figure*}

\begin{figure*}[!ht]
 \centering
    \gridline{\fig{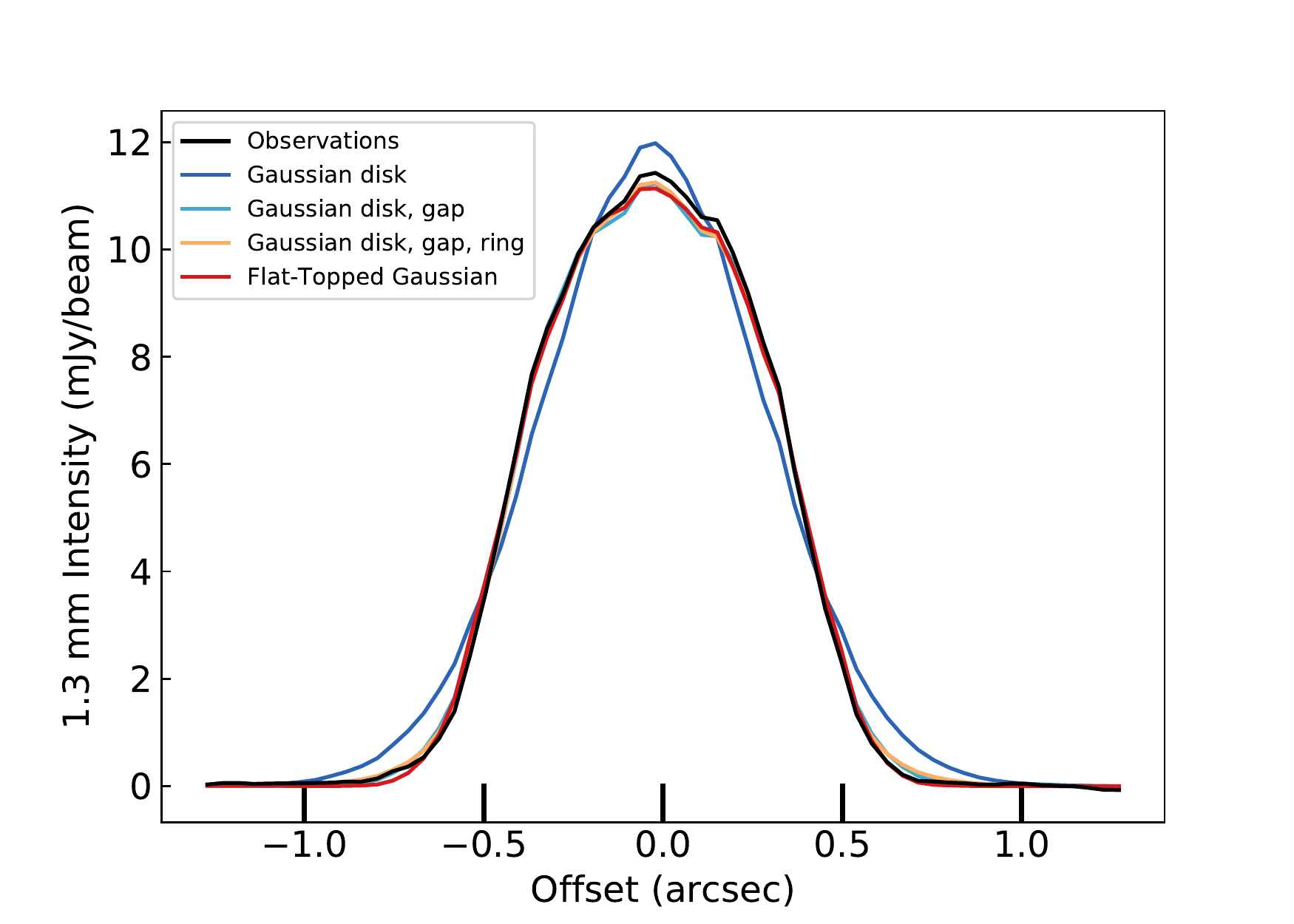}{0.5\textwidth}{}\fig{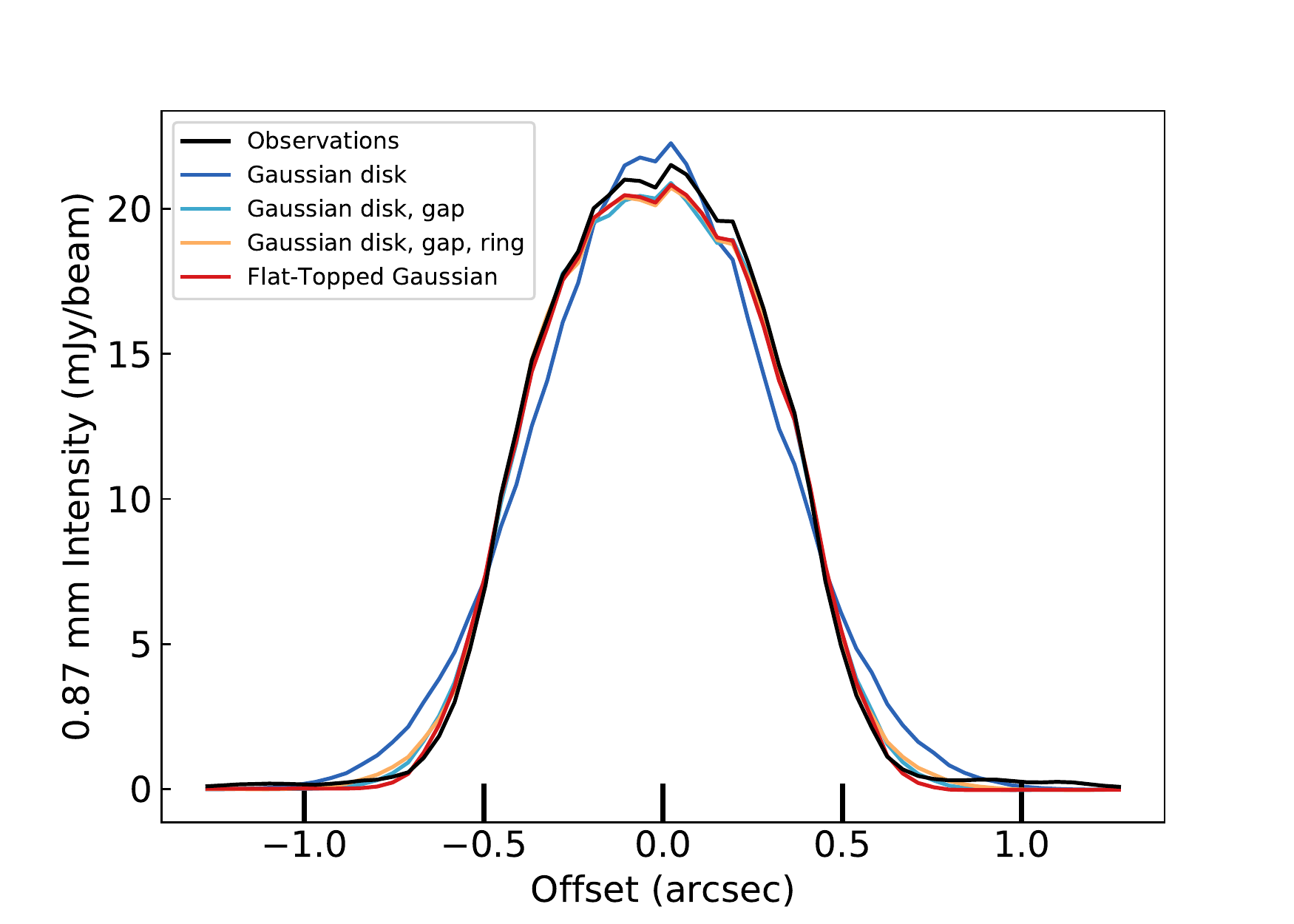}{0.5\textwidth}{}}
    \vspace{-5mm} 
    \caption{Intensity profiles along the disk major axis of the best-fit disk model images presented in Figure~\ref{fig:convolved_images}. The best-fit disk profiles prior to the \textit{uv} plane sampling, Fourier transformation, and beam convolution are shown in Figure~\ref{fig:modelsB7}. \textbf{Left:} At 1.3~mm and \textbf{Right:} at 0.87~mm. \label{fig:intensity}}
 \end{figure*}

 In this section, we present images of the best-fit disk models from Table~\ref{tbl:bestfits}. Figure~\ref{fig:convolved_images} show images of the disk models convolved with the beam at 1.3 and 0.87~mm, respectively. The Gaussian disk model is the only one from the selection which differs slightly qualitatively from both the observations and the other disk models. The other three models are virtually indistinguishable from each other and the observations when convolved with the beam at both wavelengths. Figure~\ref{fig:intensity} shows the radial profiles of the imaged models and demonstrates the similarity between the modeled data when convolved with the beam compared to the original unconvolved models in Figure~\ref{fig:modelsB7}. We note that the observed image at 0.87 mm does appear a bit boxier than the modeled data, however. This shape may reflect the flared nature of the disk, which is not captured in the geometrically thin modeled data.
 
 \subsection{Edge-on flared disk toy model} \label{sec:appendix_toygrid}
 
 In this section, we test the parameter space for the toy model and provide further details as to its design. Section~\ref{sec:toy} briefly describes the toy model and shows the result of one of the better models matching the observations. We do not fit the observations, but we show representative images of the produced residuals. We also note that the parameters being tested are not motivated by radiative transfer; this is a qualitative result used to provide a suggested explanation for the systematic residuals observed in Figure~\ref{fig:residuals}.
 
 We base the toy model disk on the \textit{Flat-Topped Gaussian} model and supplement this with a flaring shape. For simplicity, we model the disk with $i=90^{\circ}$ and $PA=0^{\circ}$ and generate the intensity profile according to the 0.87~mm \textit{FTG} fit:
 \begin{equation}
     I_1(R) = I_0 \text{exp}\left(-0.5\left(\frac{R}{\sigma_0}\right)^{\phi}\right),
 \end{equation}
 \noindent where $I_0=18.4$~mJy beam$^{-1}$, $\sigma_0=0.40^{\prime\prime}$, and $\phi=5.0$. To model the flaring, we use a Gaussian function whose width, $\sigma_h$ will change as $R$ increases away from the disk center such that:
 \begin{equation}
     I_2(R) = I_1(R) \text{exp}\left(-0.5\left(\frac{R}{\sigma_h(R)}\right)^{2}\right).
 \end{equation}
 
 \noindent The value of $\sigma_h(R)$ is modeled according to a vertical disk extent function described in Equation~\ref{eq:sigmah}. We provide the equation for $\sigma_h(R)$ again below for the reader:
  \begin{equation}
     \sigma_h(R) = h_0 \left(\frac{R}{R_0}\right)^{\beta} + h_{\text{off}}.
 \end{equation}
 
 With the generated disk intensity image, we simulate ALMA 0.87~mm observations using \texttt{simalma} with an identical setup as the 0.87~mm observations \citep{Harris_2018}, then we use \texttt{GALARIO} to recover the best fit \textit{FTG} model, and we evaluate the residuals. 

 We manually conduct a small parameter search to find reasonable fits that yield similar residuals to those obtained from the observations. We examine the effects of varying $h_0$, $\beta$, $h_{\text{off}}$, and $R_0$ on the resulting residuals. 
 
\begin{figure*}[!ht]
 \centering
    \hspace{-1mm}\gridline{\fig{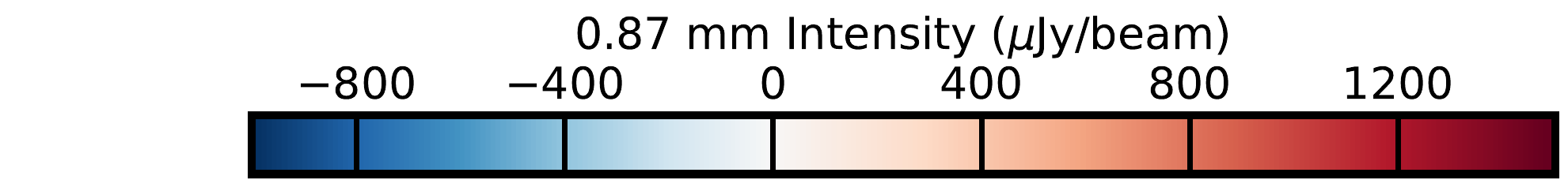}{0.65\textwidth}{}}
    \vspace{-9mm} 
    \gridline{\fig{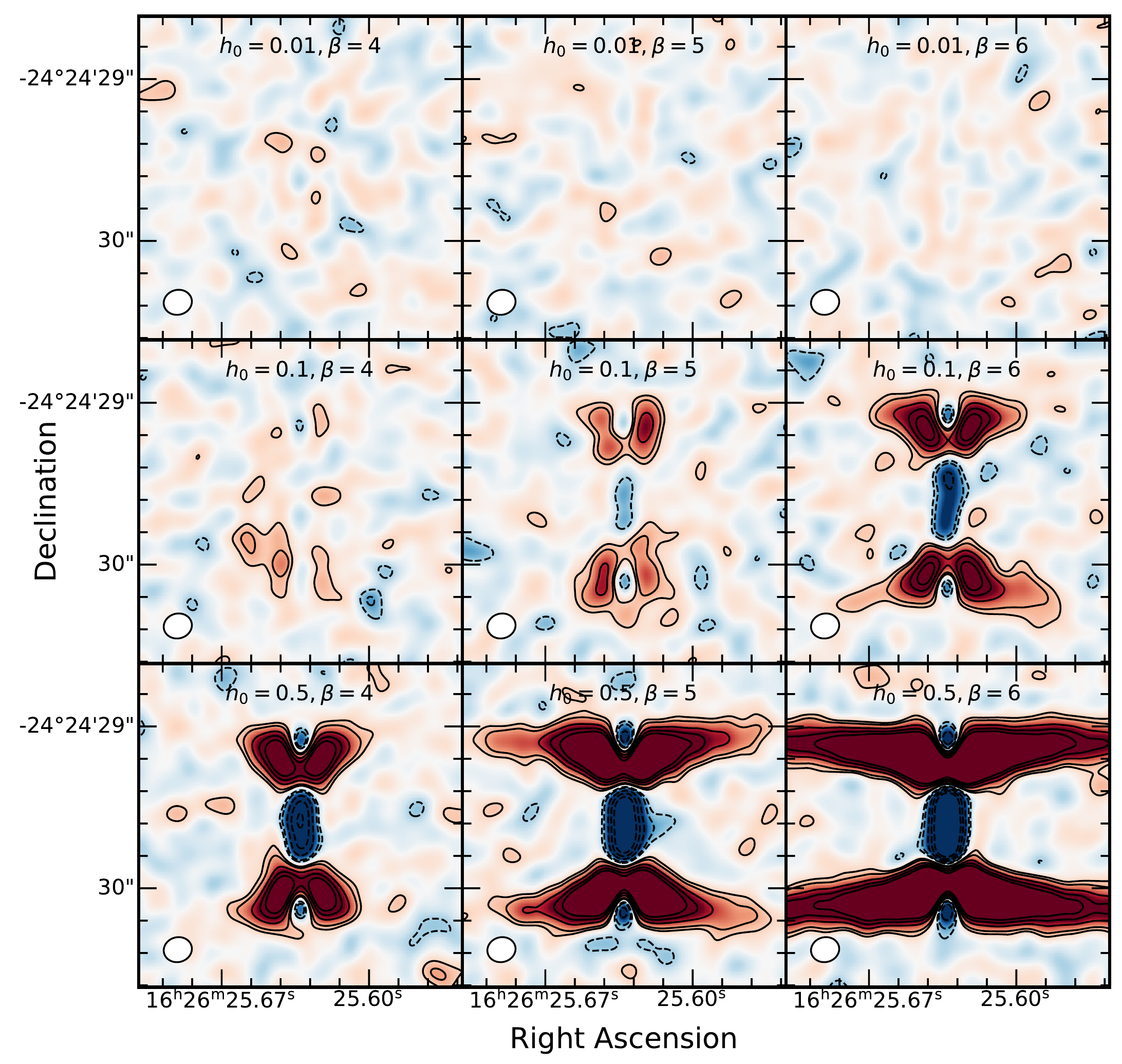}{0.98\textwidth}{}}
    \vspace{-5mm} 
    \caption{The parameters used to generate the flared disk models for this grid are $h_{\text{off}}=3$ au and $R_0=30$ au where we explore the effect of $h_0$ and $\beta$. The black contours represent $\pm 3, 5, 10, 15, 20\sigma_7$ residuals.\label{fig:grid_3AU}}
 \end{figure*}

\begin{figure*}[!ht]
 \centering
    \hspace{-1mm}\gridline{\fig{colorbar.pdf}{0.65\textwidth}{}}
    \vspace{-9mm} 
    \gridline{\fig{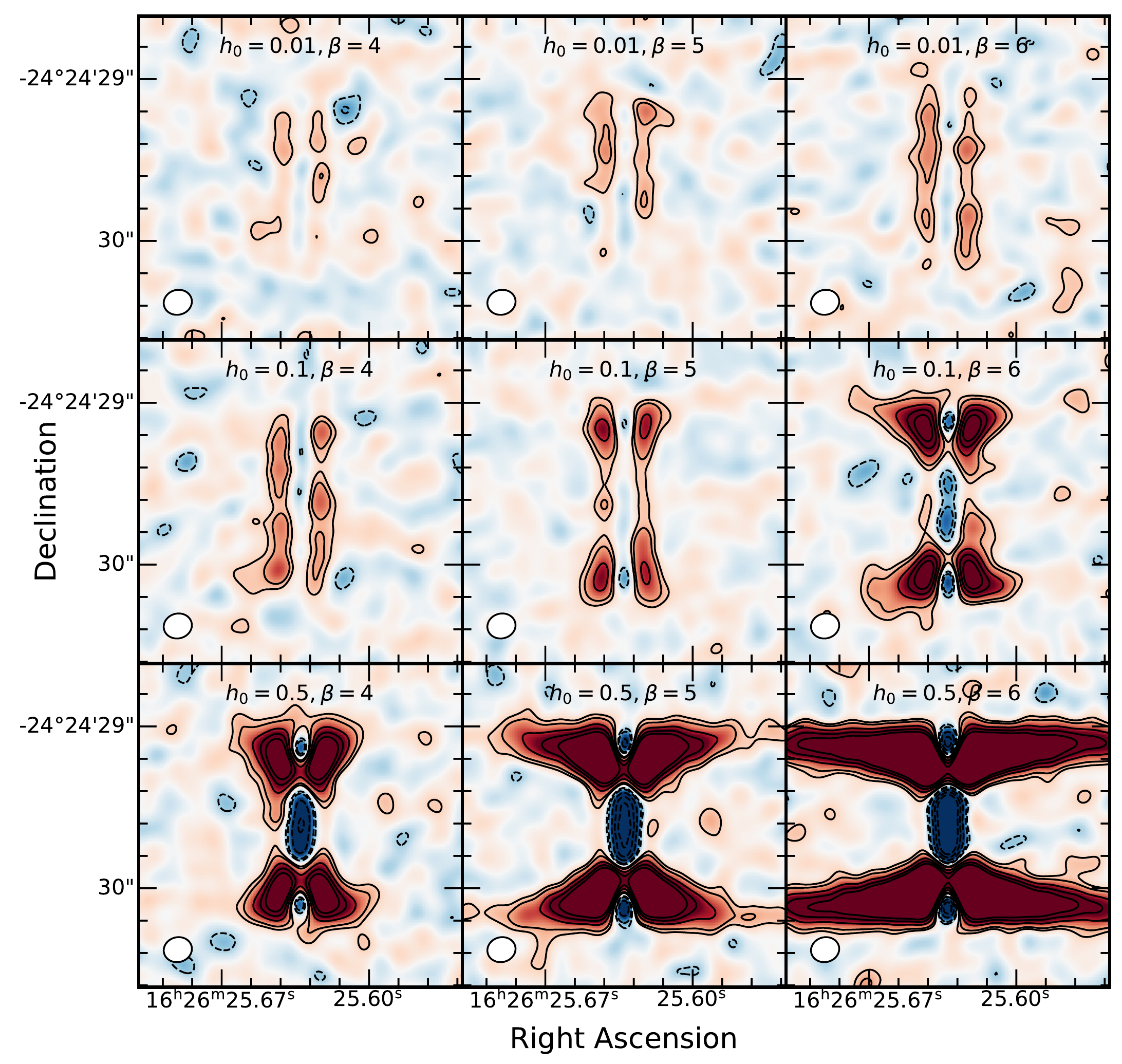}{0.98\textwidth}{}}
    \vspace{-5mm} 
    \caption{Same as Figure~\ref{fig:grid_3AU} but for this grid we use $h_{\text{off}}=4$ au. \label{fig:grid_4AU}}
 \end{figure*}

\begin{figure*}[!ht]
 \centering
    \hspace{-1mm}\gridline{\fig{colorbar.pdf}{0.55\textwidth}{}}
    \vspace{-9mm} 
    \gridline{\fig{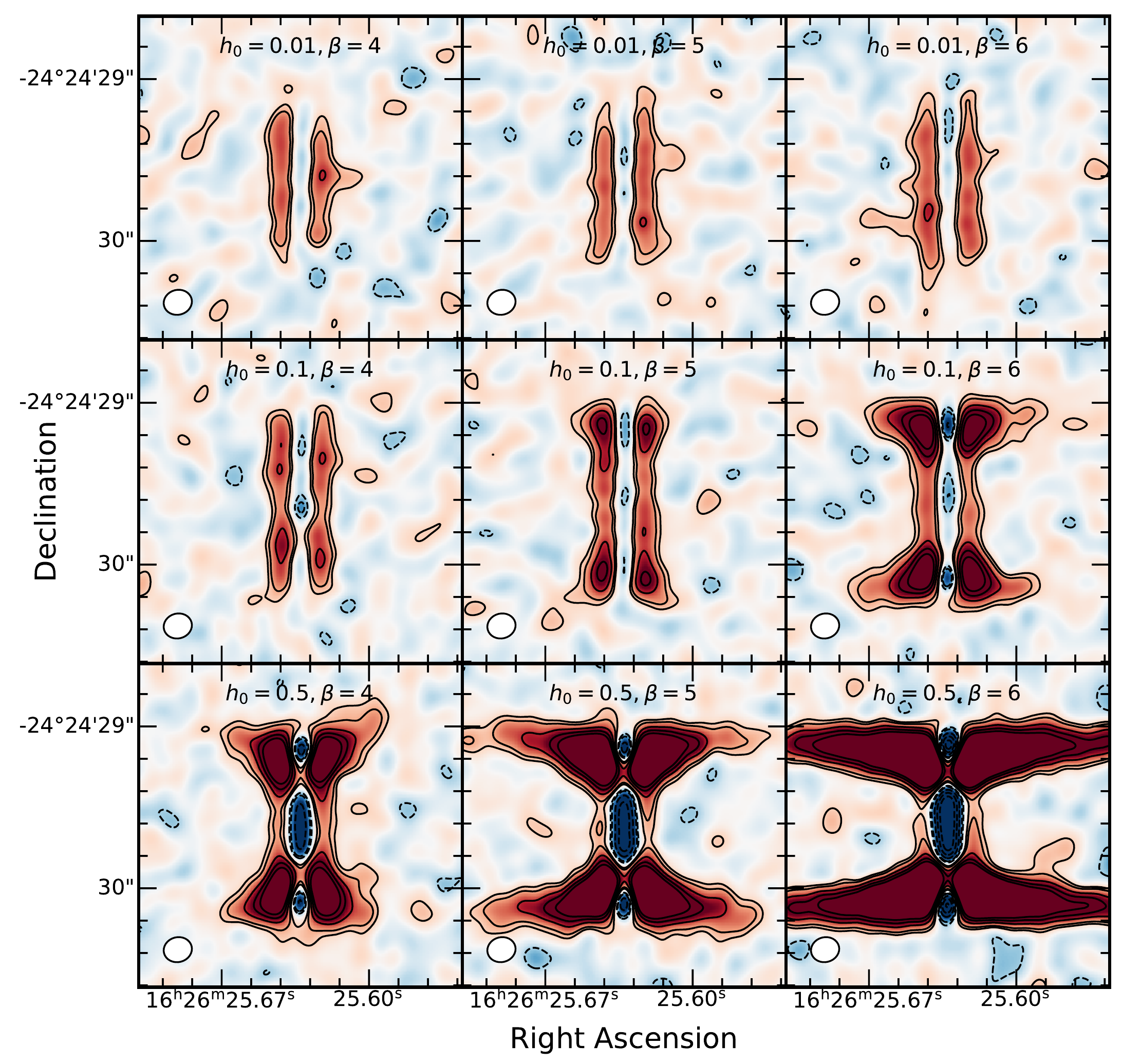}{0.98\textwidth}{}}
    \vspace{-5mm} 
    \caption{Same as Figure~\ref{fig:grid_3AU} but for this grid we use $h_{\text{off}}=5$ au. \label{fig:grid_5AU}}
 \end{figure*}

\begin{figure*}[!ht]
 \centering
    \hspace{-1mm}\gridline{\fig{colorbar.pdf}{0.5\textwidth}{}}
    \vspace{-9mm} 
    \gridline{\fig{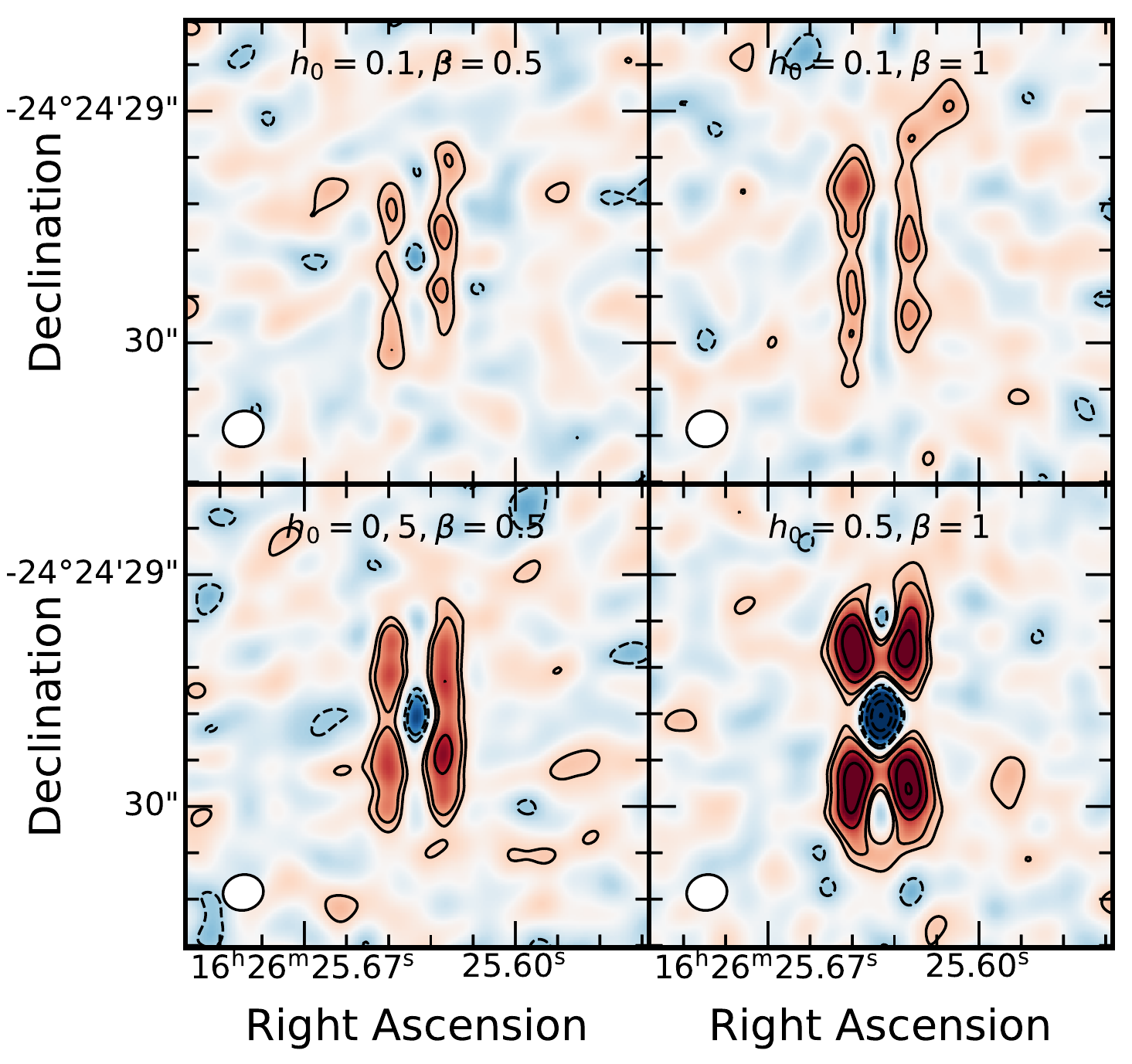}{0.6\textwidth}{}}
    \vspace{-10mm} 
    \gridline{\fig{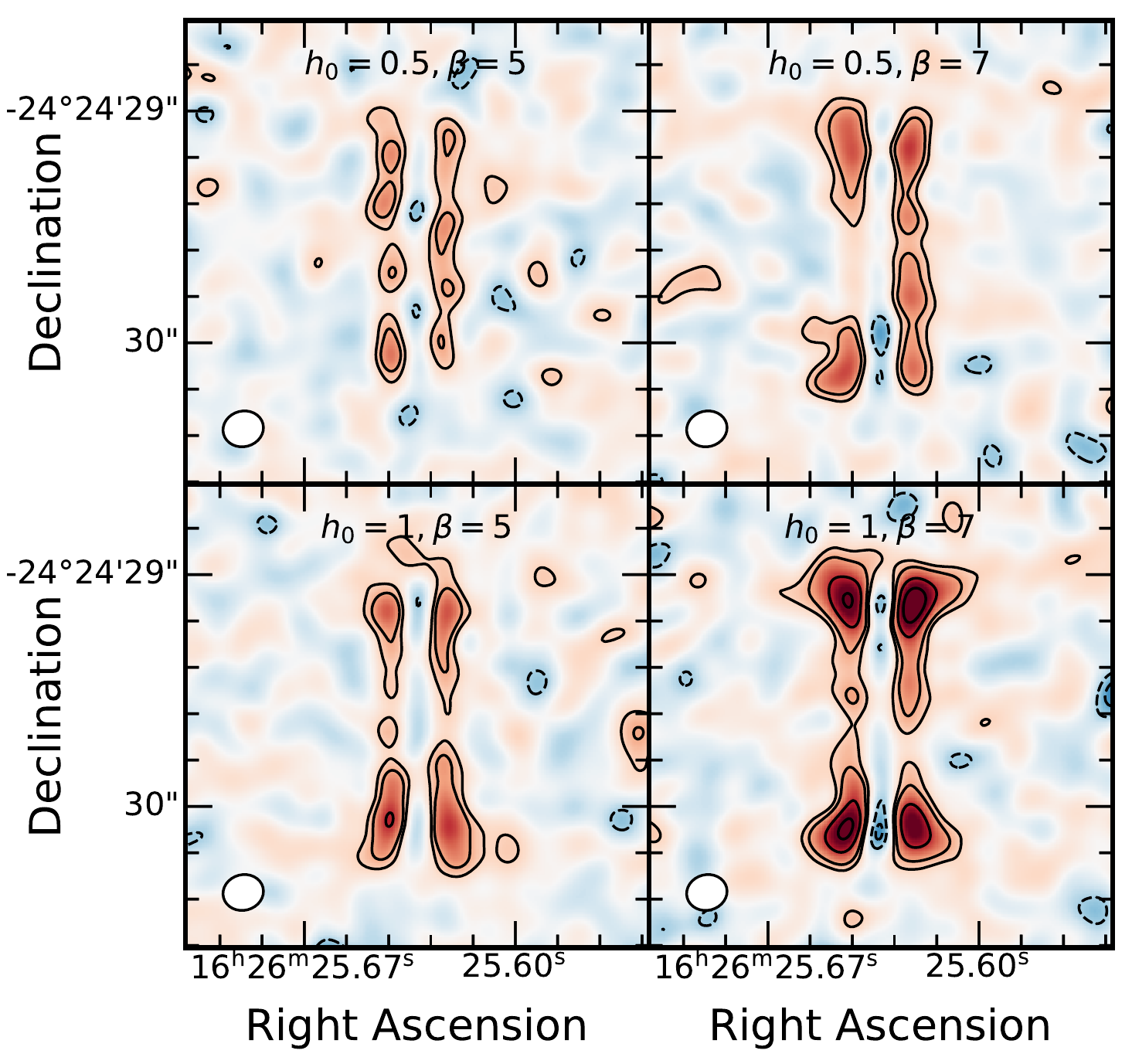}{0.6\textwidth}{}}
    \vspace{-5mm} 
    \caption{Testing the effect of changing $R_0$ on the residual morphologies. For both grids we use $h_{\text{off}}=4$ au, and the black contours represent $\pm 3, 5, 10, 15, 20\sigma_7$ residuals. \textbf{Top grid:} We use an $R_0=10$ au with varying $h_0$ and $\beta$. \textbf{Bottom grid:} We use an $R_0=50$ au with varying $h_0$ and $\beta$, which are different from $R_0$ 10 and 30 au. \label{fig:grid_4AU_R}}
 \end{figure*}

 Figures~\ref{fig:grid_3AU}-\ref{fig:grid_5AU} show the residuals for fixed $R_0 = 30$ au with $\beta=[4, 5, 6]$, $h_0=[0.01, 0.1, 0.5]$ and $h_{\text{off}}=[3, 4, 5]$ au. The top grid in Figure~\ref{fig:grid_4AU_R} shows $R_0 = 10$ au with $\beta=[0.5, 1]$, $h_0=[0.1, 0.5]$, and $h_{\text{off}} = 4$ au. The bottom grid in Figure~\ref{fig:grid_4AU_R} shows $R_0 = 50$ au with $\beta=[5, 7]$, $h_0=[0.5, 1]$, and $h_{\text{off}} = 4$ au. For the smaller grids, $h_{\text{off}}$ is fixed at 4 au motivated by the best residual from the $R_0 = 30$ au grid. The two grids at $R_0 = 10$ au and $R_0 = 50$ au are smaller to illustrate the effects of changing $R_0$, although we note that the $h_0$ and $\beta$ required different ranges from the $R_0 = 30$ au case to keep the toy model generally consistent with the observations.
 
 In general, for the $R_0 = 30$ au cases, the smallest value of $h_0 = 0.01$ produces disks with insufficient flaring resulting in residuals ($<3\sigma_7$ in most cases) that do not match the observations' residuals. The highest values of $h_0 = 0.5$ tested produce too much flaring for all cases examined. Similarly, most of the models with low $\beta = 4$ or high $\beta = 6$ produce too little or too much flaring, respectively. Thus, the best-fit cases are with $h_0 = 0.1$ and $\beta = 5$ for the grids examined in Figures~\ref{fig:grid_3AU}, \ref{fig:grid_4AU}, and~\ref{fig:grid_5AU}. We note that there are slight differences with $h_{\text{off}}$, where there is too much angled flaring for $h_{\text{off}} = 3$ au and a stronger central bridge and more box-like residuals for $h_{\text{off}} = 5$ au. Thus, we identify the best match to the observations based on visual inspection to be the $h_0 = 0.1$, $\beta = 5$, and $h_{\text{off}} = 4$ au model for $R_0 = 30$ au (see also, Figure~\ref{fig:toymodel}). 
 
 In Figure~\ref{fig:grid_4AU_R}, we show the results of adjusting the scaling radius $R_0$. For $R_0 = 10$ au (see the top grid of Figure~\ref{fig:grid_4AU_R}), the residual flaring starts too close to the inner disk region compared to the observations. For $R_0 = 50$ au (bottom grid of Figure~\ref{fig:grid_4AU_R}), the flaring is restricted to the outer disk as seen in the observations, and these toy models are comparable to those from the $R_0 = 30$ au grid. Further parameter fine-tuning will be necessary to constrain the actual disk and flaring properties. The ideal radial scaling could be between $30\leq R_0 \lesssim 50$ au with $h_0$ and $\beta$ being adjusted accordingly. There is an upper limit of $R=100$ au given that the disk radius is smaller than that. For this paper, we select $R_0=30$ au as a good approximation for the radial scaling factor.

\clearpage
\clearpage
\clearpage

\subsection{Testing \texttt{frank}}\label{sec:appendix_frank}
 
 We run \texttt{frank} from \citet{Jennings_2020} for both 50 k$\lambda$ cut data-sets providing the source coordinates, inclination, and position angles from the \texttt{GALARIO}-inferred \textit{Flat-Topped Gaussian} fits. In both cases the resulting inferred disk is highly structured and includes a gap and a ring feature over the underlying disk at 1.3~mm and possibly two rings and a gap in the 0.87~mm data. This highlights the dangers of over-interpreting features found in the \textit{uv}-visibilities for highly inclined sources. It is qualitatively similar to what we found from the complex \textit{Gaussian disk, gap, ring} model with \texttt{GALARIO} which we disfavor compared to the \textit{Flat-Topped Gaussian} model. \citet{Jennings_2020} does highlight \texttt{frank}'s limitations including the challenges of modeling disks with high inclination, optical depth, or vertical structure. VLA 1623W has all of these features, so an over-interpretation of structure is not unexpected.
 
 \begin{figure*}[!h]
   \centering
   \textbf{1.3~mm data}
   \vspace{-2mm}
   \gridline{\fig{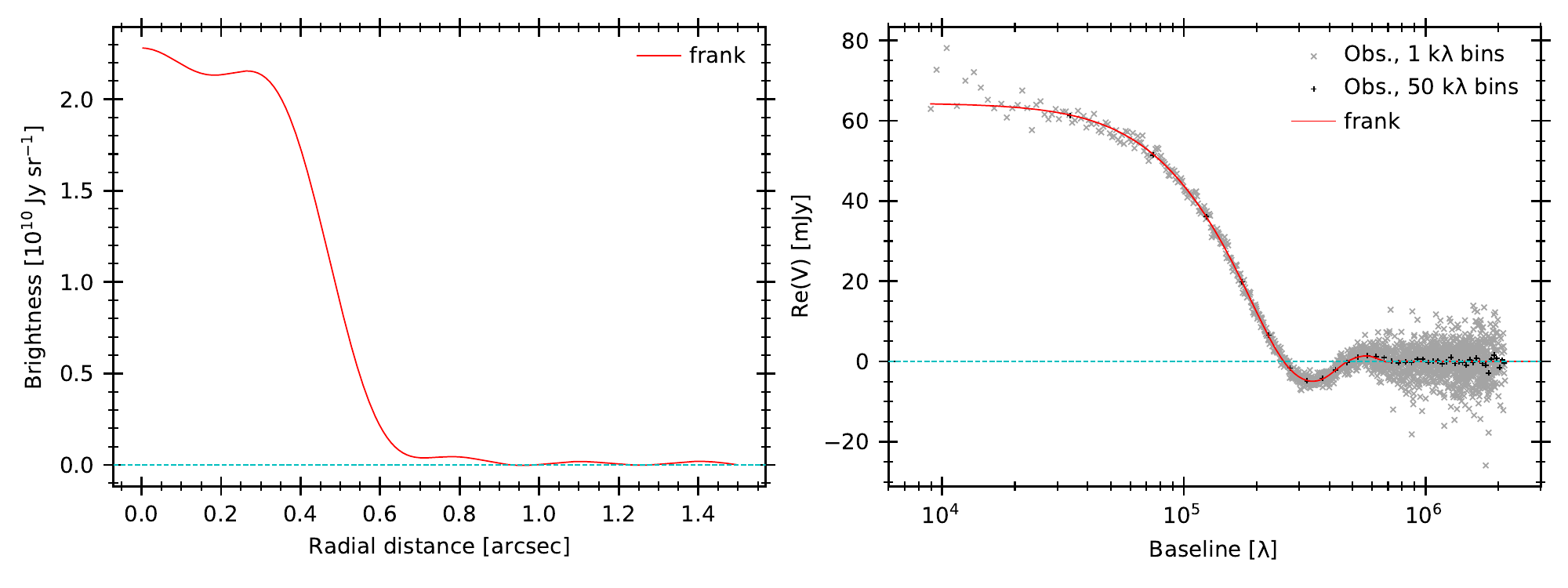}{0.95\textwidth}{}
          }
    \vspace{-5mm}
    \textbf{0.87~mm data}
    \vspace{-2mm}
    \gridline{\fig{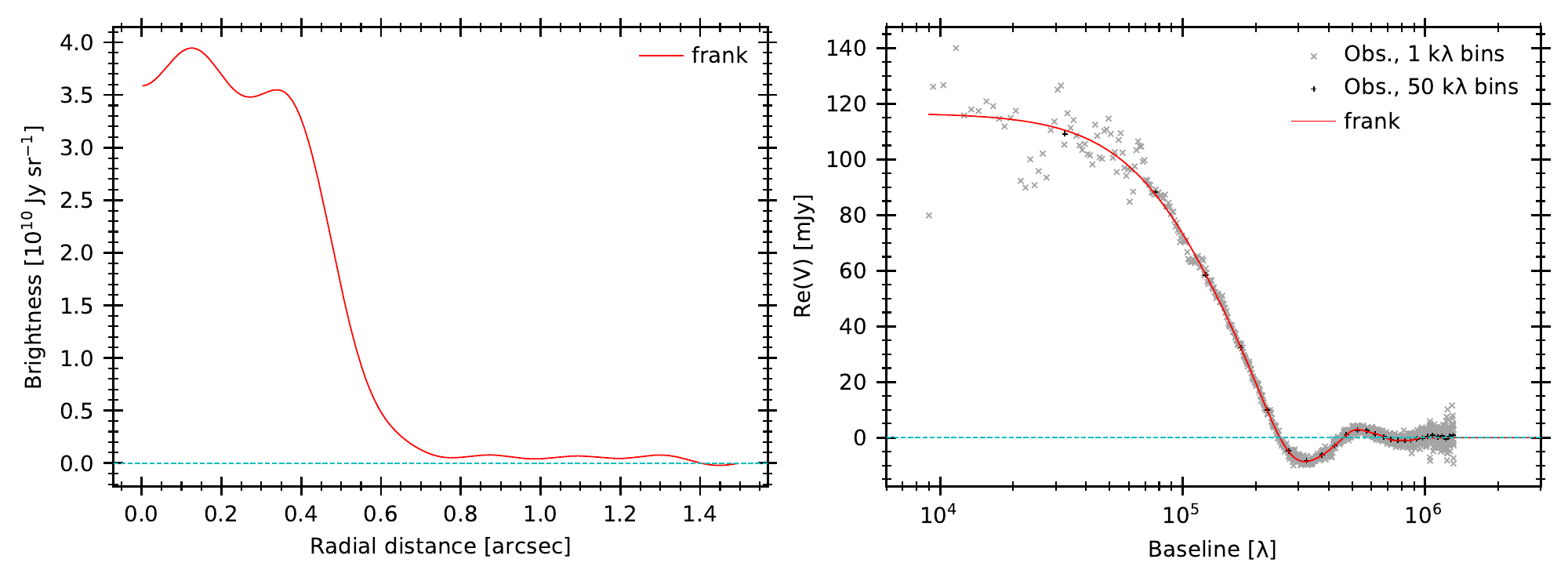}{0.95\textwidth}{}
          }
      \caption{Brightness and Real \textit{uv} profiles for \texttt{frank} fits to both data-sets. In the top row the 1.3~mm data is fit and in the bottom row the 0.87~mm data }
         \label{fig:frank}
\end{figure*}

\end{document}